\documentclass[12pt,letterpaper]{article}
\pdfoutput=1
\usepackage{graphicx}
\usepackage{subfig}

\usepackage{amsfonts,amssymb}
\newcommand{\sn}{\mathbf{sn}}
\newcommand{\cn}{\mathbf{cn}}
\newcommand{\dn}{\mathbf{dn}}
\newcommand{\am}{\mathbf{am}}

\newcommand{\sd}{\mathbf{sd}}
\newcommand{\be}{\begin{equation}}
\newcommand{\ee}{\end{equation}}
\newcommand{\ben}{\begin{displaymath}}
\newcommand{\een}{\end{displaymath}}
\newcommand{\bea}{\begin{eqnarray}}
\newcommand{\eea}{\end{eqnarray}}
\newcommand{\bean}{\begin{eqnarray*}}
\newcommand{\eean}{\end{eqnarray*}}

\def\l {\lambda}

\def\g {\gamma}

\def\s {\sigma}

\def\T {\Theta}

\renewcommand{\t}{\theta}

\newcommand{\ads}[1]{\mbox{${AdS}_{#1}$}}
\newcommand{\adss}[2]{\mbox{$AdS_{#1}\times {S}^{#2}$}}

\newcommand{\pa}{\partial}

\newcommand{\w}{\wedge}

\newcommand{\ie}{{\it i.e.}}

\newcommand{\commentout}[1]{}

\def \bS {{\bar S}}\def \fo {{1\ov 4}} 

\newcommand{\beq}{\begin{equation}}
\newcommand{\eeq}{\end{equation}}
\newcommand{\beqr}{\begin{displaymath}}
\newcommand{\eeqr}{\end{displaymath}}
\newcommand{\beqa}{\begin{eqnarray}}
\newcommand{\eeqa}{\end{eqnarray}}
\newcommand{\beqar}{\begin{eqnarray*}}
\newcommand{\eeqar}{\end{eqnarray*}}
\renewcommand{\k}{\kappa}
\newcommand{\m}{\mu}
\newcommand{\n}{\nu}

\renewcommand{\r}{\rho}

\newcommand{\cN}{{\cal N}}

\newcommand{\cO}{{\cal O}}
\newcommand{\half}{\ensuremath{\frac{1}{2}}}

\newcommand{\im}{\ensuremath{{\mathcal Im}}}

\newcommand{\bi}{\ensuremath{\bar{\imath}}}

\newcommand{\rf}[1]{(\ref{#1})}

\def \Im {{\im}}

\newcommand{\N}[1]{\ensuremath{\cN=#1}}

\newcommand{\ps}{\ensuremath{\partial_\sigma}}

\def \EllipticE  {{\rm E}} 
\def \EllipticK  {{\rm K}} 

\def \Y {{\cal X}}
 \def \AdS { \ads{}}

 \def \S {{\rm S}}
 \def \epp {\varepsilon}  \def \half {{{\textstyle {1 \ov 2} } }}

\begin{document}
%%%%%%%%%%%%%%%%%%%%%%%%%%%%%%%%%%%%%%%%%%%%%%%%%%%%%%%%%%%%%%%%%%%%%%%%
%%%%%%%%%%%%%%%%%%%%%%%%%%%%%%%%%%%%%%%%%%%%%%%%%%%%%%%%%%%%%%%%%%%%%%%%
%%%%%%%%%%%%%%%%%%%%%% TITLEPAGE %%%%%%%%%%%%%%%%%%%%%%%%%%%%%%%%%%%%%%%
%%%%%%%%%%%%%%%%%%%%%%%%%%%%%%%%%%%%%%%%%%%%%%%%%%%%%%%%%%%%%%%%%%%%%%%%
%%%%%%%%%%%%%%%%%%%%%%%%%%%%%%%%%%%%%%%%%%%%%%%%%%%%%%%%%%%%%%%%%%%%%%%%

\overfullrule=0pt
\parskip=2pt
\parindent=12pt
\headheight=0in \headsep=0in \topmargin=0in \oddsidemargin=0in
\vspace{ -3cm}
\thispagestyle{empty}
\vspace{-1cm}
\rightline{Imperial-TP-AAT-2012-08}

\today 

%\rightline{NSF-KITP-11-071}
%  somehow does not latex on my computer \rightline{\framebox[1.1\width]{\rule[-0.3cm]{}{1cm} \bf File version: \jobname    \  \ \ \ \    File compiled on: \today}}
\begin{center}
\vspace{1cm}
{\Large\bf  %Three point 
Wilson loops T-dual to Short Strings }
\vspace{1.2cm}

{M. Kruczenski$^{a,}$\footnote{markru@purdue.edu} and A.A. Tseytlin$^{b,}$\footnote{Also at Lebedev  Institute, Moscow. tseytlin@imperial.ac.uk }}\\
\vskip 0.6cm
{\em 
$^{a}$       
   Department of Physics, Purdue University,
%525 Northwestern Avenue, 
W. Lafayette, IN 47907, USA  \\
\vskip 0.08cm
\vskip 0.08cm $^{b}$ Blackett Laboratory, Imperial College,
London SW7 2AZ, U.K. 
\vskip 0.08cm
}
\vspace{.2cm}
\end{center}

\vspace{.4cm}
\begin{abstract} 
We  show that closed  string   solutions in  the bulk of  \ads{}  space are  related by 
T-duality to  solutions representing an open  string ending at the boundary of  \ads{}. 
By combining the limit in which a closed  string becomes small with a large boost,
we find that the  near-flat space  short  string  in the bulk 
%then 
maps to  a periodic open string  world surface ending on a wavy line  at the boundary.
% of the type described 
%found earlier  by Mikhailov. 
This   open string solution 
was previously found by Mikhailov and  
corresponds  to a time-like  near BPS  Wilson loop   differing  by small   
fluctuations  from a straight line. 
A simple relation is found  between  the shape of the Wilson loop and the shape of the 
closed string at the moment  when  it crosses the horizon of the Poincare patch. As a result, the energy and spin of the closed
string are encoded in properties of the Wilson loop. 
This suggests 
%It also follows 
that closed string amplitudes  with one of   the   closed strings falling  into the Poincare horizon should be dual to 
gauge theory correlators involving local operators and a Wilson loop of the T-dual (``momentum'')  theory.
% (representing the closed string that crosses the horizon).

%This  may be  suggesting  a possibility of relating   %exact in $\l$  
%coefficients in the 
%the energies of small strings and expectation values  of the  corresponding   near straight line Wilson loops.
%S14
%We comment also on potential  implications of this relation for a dual description  of a process  when 
% a short closed string crosses  a horizon. 

%This T-duality   relation may be   viewed as  a way of finding new Wilson loop solutions
%(assuming   formal decompactification of  spatial  string direction $\s$ at the end)  
%with  direct connection to the   original  closed string, 
%but our original motivation was to understand a  similarity between   Basso's  exact (in $\sqrt \lambda$) 
%slope 
%expression for short small-spin folded string  and Correa et al   exact expression for the  coefficient of 
%the leading  term  small-angle or wavy line   Wilson loop expectation value. 

\end{abstract}

\newpage
\setcounter{equation}{0} 
\setcounter{footnote}{0}
\setcounter{section}{0}

\setcounter{equation}{0} \setcounter{footnote}{0}
\setcounter{section}{0}
\vfuzz2pt % Don't report over-full v-boxes if over-edge is small
\hfuzz2pt % Don't report over-full h-boxes if over-edge is small
\setcounter{footnote}{0}

%%%%%%%%%%%%%%%%%%%%%%%%%%%%SEC1%%%%%%%%%%%%%%%%%%%%%%%%%%%%%%%%%%%%%%%%%%%%%%%%
\def \no {\nonumber}
 
%\keywords{Classical string solutions, AdS/CFT, Wilson loops}

%\preprint{\tt{} \\
%          \tt{hep-th/yymmnnn}  }
\def \eps {\epsilon} \def \la{\label}  \def \ci {\cite} \def \be {\beqa}  \def \ee {\eeqa}
\def \X {{\cal X}}  \def \m {\mu}  \def \n {\nu}
\def  \mae  {\mathbb E}
\def  \maf  {\mathbb F}
\def  \mak  {\mathbb K}
\def \lab {\label}
\def \foot {\footnote}\def \del {\partial}
\def \td {\tilde}
\def \t {\tau}  \def \s {\sigma}
\def \ep {\epsilon}   \def \ov {\over}
 \def \sql {\sqrt{\lambda}}
\def \ep {\epsilon}
\def \t {\tau} \def \s {\sigma}  \def  \rx   {{\rm x}} \def \ha { {1 \ov 2}}

\def \T {{\cal T}} 
\def \ry {{\rm y}}

 \def \ha {{1\ov 2}}

\def \AdS {{$AdS\ $}}
\def \ed {\end{document}}
%%%%%%%%%%%%%%%%%%%%%%%%%%%%%%%%%%%%%%
\tableofcontents

\renewcommand{\theequation}{1.\arabic{equation}}
\setcounter{equation}{0}

\section{Introduction}

There are many   known relations between   closed and open strings. 
For example, in flat space  a  world sheet  representing a propagation of a  closed string  may be viewed, after $\t \leftrightarrow \s$
interchange,     as describing  also 
propagation  of an open string in periodic time; an open string   disc  diagram  for open strings may be  viewed as a closed   string 
emission by a D-brane into vacuum; KLT relations  express closed string scattering  amplitudes in terms of open  string scattering amplitudes, etc.

In the gauge-string duality context,  one recent example  is the relation between   the  coefficient 
of  the leading logarithmic term in the {\it large}  spin  (long string) asymptotics of the  closed   folded spinning string energy \ci{gkp} 
and the  cusp anomalous   dimension  of the Wilson loop defined by a  light-like  cusp \ci{mk,kt,sev}. 
There   is also  a (KLT-like)   relation between  a correlator   of  closed string   vertex operators  at null separations 
and a   square   of a  Wilson loop  defined by the corresponding null polygon  \ci{all}.

These are  cases  of   far-from-BPS configurations,   but there are other   examples  that  hint towards  a possible closed-open string    
relation  also  for  ``short''  strings or  ``small''  deviations    from the   BPS limit. 
According to \ci{cor},    for a  straight line Wilson loop  with a small cusp of angle $\pi -\phi$,  $ \phi \ll 1$   in $AdS_5$
one finds\foot{Here we  quote  the coefficient function in  the  planar approximation only ($\l$ is the  't Hooft coupling). 
  $\Lambda$   is a UV  cutoff   and $I_n$ is a modified Bessel function.} 
\be 
&&  \langle W_{\rm cusp}  \rangle = \exp \Big[ - \Gamma_{\rm cusp} (\phi, \l)   \ln {   \Lambda  } +... \Big]  \ , \la{1}\\ 
 &&
 \Gamma_{\rm cusp} (\phi, \l)_{_{\phi \to 0}}     = - B(\l) \, \phi^2 +  ... \ , \  \la{2}  \\
 &&    B(\l) = { \sql \ov 4 \pi^2 } {  I_2 (\sql) \ov I_1 (\sql) }      %=\Big|_{\l \gg 1}  
  =  { 1 \ov 4 \pi^2} \Big( \sql   - {3\ov 2}    + {3 \ov 8 \sql} + ...  \Big)   \  .      \la{3} 
 \ee
 In \rf{3}  we gave the  expansion of the coefficient  $B(\l) $  at strong coupling or 
 for large  string tension, i.e.  $\l\gg 1 $.\foot{The coefficient 
 %25
 $B$ is also proportional to the correlator of  the circular Wilson loop \ci{sz}  with the  integrated  dilaton operator, i.e. 
 $B = { \sql \ov 4\pi^2} { d \ov d \sql} \log \langle   W_{\rm circle} \rangle$.}
 The same expression   is found for a  cusp of angle $\pi -\theta$  in $S^5$ (with $\phi^2 \to - \theta^2$). 
 It  was observed  in \ci{gs}  that  switching on   an   orbital momentum $J$ in $S^5$,   the expression for the leading $\l \gg 1$  term in 
 the corresponding $\Gamma_{\rm cusp} (\theta, \l; J) $    appears   to be related
to  the   small-spin  limit of  the energy of a  folded spinning string in $S^3$. 
 
 \def \S {{\rm S}}
 
 Moreover,  there  is a  striking similarity \ci{cor}   between the   {\it small}  angle  coefficient  $B$ in \rf{2}      and the slope   function  \ci{bas}  
 found in the   {\it small}  spin limit of the  $AdS_5$ folded string energy. 
 According to   \ci{bas,gros,bas2},   
the   dimension   of  the  sl(2)  sector  gauge-theory operator    with 
 spin $\S $ and   twist $J$   or the energy of the corresponding  dual   spinning string state 
  has the following  expansion  in the formal  $S \ll 1$ limit\foot{This result follows  directly  from the asymptotic Bethe ansatz 
  expression for the string spectrum. It  is not   sensitive to the dressing phase,   which  should have to do with its 
  ``near-BPS''   nature    avoiding non-trivial   order-of-limits or renormalization issues, see \ci{bet}.  Note that  the    computability 
  of the  coefficient $B$ in \rf{2} is, in turn,  due to  supersymmetry allowing one  to use localization techniques to compute  certain correlators 
  of  BPS Wilson loop  with   particular  local operators.}
\be 
&&E^2 =   J^2 +  h (  \l,J)\,   \S +   O(\S^2) \ , \la{4} \\
&&  h= 2 J +   \bar h (\l, J)\ , \ \ \ \ \ \ \ \ \    \bar h (\sql, J) = 2    \sql\,   {  I_{J+1} (\sql) \ov I_J (\sql) }   \ .  \la{5} \ee
If one   formally  sets  here $J=1$
  then  the  slope function  in \rf{5}   becomes    directly related   to $B$   in eq.\rf{3}, i.e.\foot{Heuristically, the  $J=1$  choice  
  may  be thought of    as  required to   represent
 an open string ``half'' of a closed string state  dual to sl(2)   sector operator  with the minimal possible value of $J=2$.}  
   \be 
 % \bar h  (\sql, 1) =   8 \pi^2  B =   2 \sql   - {3}    + ...    \ , \ \ \ \ \ \ \ \ 
  h  (\l, 1) =  2 +   8 \pi^2  B (\l)  =   2 \sql   - {1}     +   O( { 1 \ov \sql})   \ . 
  \la{6}  \ee 
%This suggests   relation  for $J=1$. Why $J=1$? Heuristically,  this  may be  related to the fact that the lowest   closed string state 
%has $J=2$   and is dual to closed  folded spinning string; cutting  string  in half to get an open one  would give   a $J=1$ state. 
A motivation    behind the present work is to try to find 
a possible relation between    small (nearly point-like) closed strings in $AdS_5$     and  long  open strings  ending at the boundary 
 (and thus corresponding to nearly-straight Wilson  lines).

Let us  recall   that the    coefficient $B(\l)$  in \rf{2},\rf{3}    has also  other interpretations. 
  As was argued in \ci{cor}, 
   for a {\it euclidean }  Wilson loop    defined by    a curve   which  is a  small  transverse  deviation from  a  straight line  \ci{poly,sem}  one has 
\beq
  \langle W_{\rm wavy}  \rangle = 1 + \half B(\lambda) \int d\sigma d\sigma' \ \frac{\left[\partial_\sigma \vec x_\perp (\sigma)-\partial_\sigma \vec  x_\perp(\sigma')\right]^2}{(\sigma-\sigma')^2} + ... \ , \la{7}
\eeq  
 where it is assumed   that   $|\partial_\sigma \vec x_\perp|\ll 1$. Furthermore, 
 the same     $B(\l)$   appears also in 
 %The function $B(\lambda,N)$ is known exactly  and at  large $N$ is given by
% \ci{cor}   \beq
%   B(\l, N)  = B(\l)  +  \cO(\frac{1}{N^2}) \ , \ \ \ \ \ \ \ \ \ 
 %   B_(\l) =   \frac{1}{4\pi^2} \frac{\sqrt{\lambda}\, I_2(\sqrt{\lambda})}{I_1(\sqrt{\lambda})}
 %   \ \to_{_{\sql \gg 1}}  \    { \sql \ov 4 \pi^2 } + O(1) \la{234}  \eeq
% The same  function  appears \ci{cor} also  in  
 the expression for the energy emitted    \ci{mik}  by a  slowly moving  ``quark''  \ci{cor} (see also \ci{fi})
 \be 
 E=  2 \pi B(\lambda)   \int  d\t  \   {\dot  v} ^2  + ... \ , \ \ \ \ \ \ \ \ \ \ \ \   \vec v = \dot {\vec   \rx} (\t)\   \ll 1   \ , \la{vv}
 \ee
 where $ \vec x  (\t)$  is  a spatial  deviation of the quark's trajectory from  a  straight line. 
 As was   shown   in  \ci{mik}, the {\it Minkowski}  open-string  world surface ending  on   the  quark's trajectory 
 at the boundary   of the Poincare patch  %s implicitly determined by   specifying  ${\vec   \rx (\t) }$
  leads to the  following expression for  the classical   ($\sql \gg 1$)  string  energy  %corresponding to $x_0$ time: 
 \be 
 E_0 = { \sql  \ov 2 \pi} \int d\t \  { \dot  v ^2 -   [\vec v \times \dot {\vec v } ]^2\ov  (1-    v ^2)^3 } 
\approx   { \sql  \ov 2 \pi} \int d\t \   \dot  v ^2    + ... \ ,  \la{er} 
  \ee
 which  is a  relativistic generalization   of  the integral    in  \rf{vv}.

The derivation  of the  classical  energy  formula \rf{er}  is straightforward  in the limit of small    deviations   from a  straight line \ci{mik}.
Using the Poincare patch coordinates 
$ ds^2 = { 1 \ov z^2} (   dz^2 - d x_0^2  + dx_i^2)$
 and choosing 
 the static gauge 
\be x_0=\tau,\ \ \ \ \     \  \ \ \ \ \    z= \s\ ,  \la{111}   \ee
  the  string (Nambu)  action for small fluctuations   in  $x_i$-directions    %near the straight  line along $x_0$  
  is 
 %in the Poincare patch $ds^2 =  z^{-2} ( - dx_0^2 +   dx_i^2 +    dz^2) $ is 
 \be 
 S= {\sql \ov 2 \pi} \int { d \t d \s  \ov  \s^{2}}  \Big[1 +    \ha    ( \dot x_i^2 - x_i'^2)  + O(\del x^3)\Big]  \ . \la{sy} 
 \ee 
The corresponding equation
 of motion   $ \ddot x_i - x''_i   + 2  \s^{-1}   x'_i=0$ 
has solution  $x_i(\t,\s)$ 
 which is  uniquely determined by the {\it two}   boundary data functions\foot{This is a Minkowski, not a Euclidean  Wilson line  stationary 
 surface; it is  the latter  which is 
(at least locally)    uniquely determined  by the boundary curve.}
 --  $\rx^+(\t)$ and $\rx^-(\t)$, or, equivalently,  
       the shape of the boundary  curve $\rx_i(\tau)\equiv  x_i(\tau, 0)=\rx^+ (\t)  + \rx^- (\t) $ 
 {\it and}    its 3-rd  normal derivative  \ci{mik}. Explicitly,  one finds\foot{ Note that 
 the near-boundary (small $\s$) expansion is 
$ x_i (\t, \s) =  \rx_i(\tau)   - \ha  \s^2 \ddot   \rx_i (\tau) +   O(\s^3) $  making it   clear that the third $\s$-derivative is an independent   boundary data.}
\be 
 x_i (\t, \s) &=&\rx_i(\t,\s)  - \s \del_\s \rx_i(\t,\s) \no \\ 
&=&
   \rx^+_i(\tau + \s)  + \rx^-_i(\tau - \s) 
  - \s \Big[ \dot {\rx}^+_i (\t + \s)   -   \dot {\rx}^-_i (\t - \s) \Big] + O( \rx^2) \ , \la{22}\\
x_i (\t, 0)  &\equiv& \rx_i(\tau)  = \rx^+_i(\tau)  + \rx^-_i(\tau ) \ , \ \ \ \ \ \ \ \ \
\del_\s^3 x_i (\t, 0) = - 2 \del_\tau^3  [ \rx^+_i(\tau)  -  \rx^-_i(\tau ) ]  \ , \la{23}
\ee
where  $ \rx_i(\t,\s) =  \rx^+_i(\tau + \s)  + \rx^-_i(\tau - \s) $   is generic  flat-space solution (harmonic function). 
%The reason that the third derivative is needed is that the near-boundary (small $\s$) expansion is 
%$ x_i (\t, \s) =  \rx_i(\tau)   - \ha  \s^2 \ddot   \rx_i (\tau) +   O(\s^3) $.
The string energy  associated with  translations in the    $x_0=\tau $   direction  
\be 
E_0 = { \sql \ov 2 \pi} \int^\infty_0   {d \s \ov \s^2} \Big[ 1 + \ha    ( \dot x_i^2 + x_i'^2)   +  O(\del x^3)\Big]  \ , \ \ \ \ \la{iy} 
 \ee 
then   leads, after renaming the integration variable 
$\s \to \t$,  to  the same expression as in \rf{er}, i.e. 
$ { \sql \ov 2 \pi}  \int  d\tau    \  { \ddot \rx}_i ^2$.\foot{The   derivation requires 
 regularizing near the boundary 
 $z=\s=\ep\to 0 $      and dropping the singular free-motion and  boundary terms, see section 3.3 below for more details.} 
%One needs  to notice that energy is 
%conserved in time so that  we can choose $\tau=0$   and then rename $\s$ into $\t$.

In this paper   we   will   show    that  T-duality  along the boundary  directions of  $AdS_5$    
relates the world sheets   of    small   closed strings   in the bulk of \ads{}   to  the open string surfaces 
ending on  wavy lines   representing  small-velocity   ``quark''  trajectories  at the boundary. 
 Our  starting   point   will be   the  observation  that    T-duality   along the  boundary directions 
  in the Poincare patch  of \ads{}   space maps  a massless 
  geodesic in the bulk of \ads{} into a  straight line  Wilson loop surface. 

As was  first  observed   in \ci{kts}, the formal T-duality along all $x^\mu= (x^0, x^i)$  boundary 
directions   maps \AdS into \AdS  space 
 provided   it  is  combined with  a   simple   coordinate transformation --  the inversion of  the radial direction, 
$z \to 1/z$
 (which effectively interchanges  the boundary and the horizon). 
 
 This  T-duality  was used  (in Euclidean  world sheet  case)    in \ci{am} 
   to relate  (imaginary) world sheet solution  that dominate a   semiclassical path integral   for  open-string (gluon) scattering  amplitudes 
   to (real)   solutions 
    describing the corresponding  null polygon  Wilson loops  in the dual momentum space. 
  %Here we shall   present another application of this   T-duality
 Here   we  will   be  considering   the   case of the {\it Minkowski}   signature in both  the  target space and   the world sheet,   so  that 
 the T-duality  transformations will  be mapping  real solutions into real solutions.

 We shall start  in section 2    with a  review of the basic   definitions 
 and properties  of T-duality transformation  in \ads{}  and   discuss some simple examples.
 % and the form of these transformations   with some examples in section 2.

 In section 3 we shall  describe  in detail   the  map   between small 
 closed strings and  long open strings  ending 
 on a nearly straight  line at the boundary of  \ads{}  and thus corresponding to the wavy line Wilson loops. 
 %closed strings and   open string surfaces ending on wavy  lines  at the boundary. 
 %{\bf expand ?} 
 
 %AT1
 In section 4 we shall   consider  an  example of  application of  the T-duality to a closed string of finite size --   
 folded  string  with  an arbitrary spin.   %solution. 
 Section 5    will contain   some   comments  on a  physical  interpretation of the T-duality relation
 and implications for a process of a  small closed string crossing a horizon. 
% {\bf expand ?} 
 Some concluding remarks will  be made in section 6.
 
   In   Appendix A we shall   give our definitions  of elliptic functions used  in section 4.
   In Appendix B we shall discuss the application of the T-duality transformation to the  small spiky string solution.  
 Appendix  C  contains a  discussion of implementation of T-duality on a cylinder  a closed string insertion.

 %  for a string in  \ads{5}  combined with   $Z \to 1/Z$   may be written  as 
 %$\del_a \td \X^\m = \ep_{ab}  Z^{-2} \partial^b \X^\m $  (with $\ep_{\t\s} =-1$),  i.e. 
%\be \la{tt}
%\del_\
%Our discussion below will apply to {\it Minkowski}  signature case when 
%T-duality will map a real string solution to a real string solution. 
%We show that   it relates \rf{22}  to small   string solution in the middle of AdS. 

%T-duality  formally relates classical string solutions to  solutions and maps  the corresponding sets of conserved charges.
% For example,  world surfaces of  massless geodesic in AdS is  T-dual to  straight-line  Wilson loop surface  (in Minkowski world-sheet signature) -- 
 %apparently not a widely known fact. 

%%%%%%%%%%%%%%%%%%%%%%%%%%%%%%%%%%%%
\renewcommand{\theequation}{2.\arabic{equation}}

\setcounter{equation}{0}

\section{Definition of   T-duality transformation   and some examples}

%{\bf replace $T$, $X$      by $\T$, $\X$ ?}

Let us start with describing    the   basic sets of coordinates in   \ads{5}   we will be using. 
The  embedding coordinates are  defined by 
\beq
 ds^2_{_{AdS_5} }=   dX_M dX^M\ , \ \ \ \ \ \ \ \ \ \    - X_M  X^M= X_{-1}^2 + X_0^2 - X_1^2 - X_2^2 - X_3 ^2 -X_4 ^2 =1\ .    \la{ep}
\eeq
It is convenient to introduce also the light-like coordinates 
\beq
X_\pm=X_{-1} \pm X_4  \ ,  %\ \ \Rightarrow 
\ \ \  \ \ \ \ \ \ 
X_+X_- -  X_\m  X^\m  %- X_1^2 - X_2^2 - X_3 ^2 
= 1\ , \ \ \ \ \ \  \ \ \ \ \     X_\m = (X_0, X_i) \ .  \la{2.2}
\eeq
The  global coordinates %a1
$(t,\rho,\hat{n}_r)$ 
are related to $X_M$ by 
\be
  &&X_{0} + i X_{-1} = \cosh\rho\ e^{it}, \ \ \ \ \ \ \ \ \  X_{r} = \sinh\rho\ \hat{n}_r, \ \ \ \ \ \ r=1,\ldots, 4 \ , \la{nnk} \\
  &&  ds^2 = - \cosh^2 \rho\ dt^2 + d \rho^2 + \sinh^2 \rho\ d \Omega^2_{[3]}\ , 
   \lab{nk}
\ee
where $\hat{n}_r$ is a 4-component  unit vector parametrising an $S^3$. 
The Poincare coordinates  are defined  by\foot{In contrast to Introduction,  in  what follows we shall use capital $Z$ and $\X^\m$ to denote the Poincare patch coordinates.
Let us recall  that the  Poincare coordinates 
only cover part of the Minkowski \ads{5} space, i.e.  there is a choice of the patch they cover. The one used here 
is appropriate for the solution and limiting procedure described below but other simply related options are possible too.
}
\be
 &&Z = \frac{1}{X_-} \ , \ \ \ \ \ \ \ \ \ \ 
 X_- =  X_{-1} - X_4 =  \cosh\rho\, \sin t -\sinh\rho\,  \hat{n}_4 \ , \la{p1}\\
 &&
   \X_0 \equiv \T=    \frac{X_0  }{X_-} =   \frac{ \cosh\r\,  \cos t        }{X_- }\ , \ \ \ \  \la{p} \\
  &&
  \X_i = \frac{X_i}{X_-}   =   \frac{ \sinh\r\ \hat{n}_i        }{X_- }\ ,\ \ \ \ \  \ \  \ i=1,2,3\ , \la{p2} \\
 &&  ds^2 = {1 \over Z^2}   ( d Z^2  +  d \X_\m d\X^\m)   \ , \ \ \ \ \ \  \ \ \ \ \   \X_\m  = (\T, \X_i), \ \ \ \ \ \  { \m=0,1,2,3}.      \lab{pok} 
\ee
We   shall   use the conformal gauge in which   the   classical     bosonic 
string action $S= {\sql \ov 2 \pi} \bS$,  equations  of motion 
and the conformal-gauge constraints   written  in the embedding coordinates  are
%($\del_\pm = \del_\t \pm \del_\s$) 
%() % \ $X_\mu  X^\mu   = - X_0^2 + X_i^2$) 
\be
&&  \bS = \ha \int d\s d\t  \left(\pa_a X_+ \pa^a X_-   -   \pa_aX_\mu \pa^a X^\mu \right)
     + \ha \int d\s d\t \Lambda \left( X_+ X_- -  X_\mu X^\mu  -1\right) \ , \la{ca}\\
&&  
 \pa_a   \pa^a X_\pm   = \Lambda X_\pm \ ,  \ \ \ \ \ \ \ \ \ \ \ \   \pa_a \pa^a X^\mu = \Lambda X^\mu 
\ ,  \cr
 && \Lambda = \del_a X_+ \del^a X_-   -  \del_a X_\m \del^a X^\m   \ ,    \ \ \ \ \ \    \ \ \  X_+ X_- - X_\m X^\m  = 1 \ ,  \la{bb1}
\\
&& \pa_\s X_+ \pa_\s X_- +\pa_\t X_+ \pa_\t X_- -    \pa_\s X_\mu \pa_\s X^\mu -  \pa_\t X_\mu \pa_\t X^\mu =0\ ,\no  \\
&&  \pa_\s X_+ \pa_\t X_- +\pa_\t X_+ \pa_\s X_- -  2\pa_\s X_\mu \pa_\t X^\mu =0 \ . \la{bb2} 
\ee 
%\be \la{e}
%&&\del_a \del^a   X_M   + \Lambda  X_M = 0 \ , \ \ \ \ \ \    \Lambda = - \del_a X_M \del^a X^M \ , \ \ \ \      X_M X^M =-1 \ , \\
%&&
%\del_+ X_M \del_+ X^M = 0\ , \ \ \ \ \ \ \ \ \del_- X_M \del_- X^M  =0 \ . \la{vii} 
%\ee
 In what follows we shall  %consider  solutions  of these   equations 
  map  one class of  solutions of  these equations  (small closed strings) into 
 another  class of  their     solutions (open-string Wilson-loop type solutions) using the formal T-duality  (2-d  scalar-scalar duality) 
 transformation    that maps \AdS into \AdS  if   combined with the $Z \to 1/Z$ coordinate transformation 
  %once combined with a coordinate transformation $Z \to 1/Z$
% (that interchanges the boundary and the horizon) 
  \ci{kts}. For brevity, we shall  often  refer to this combined    transformation simply as T-duality. 
  
  Using the Minkowski   signature in both the target space and the world sheet 
 the T-duality  transformation rule  is then $(Z,\X_\m) \to (\td Z,\td\X_\m)$ 
 where
 %  for a string in  \ads{5}  combined with   $Z \to 1/Z$   may be written  as 
$\del_a \td \X_\m = \ep_{ab}  Z^{-2} \partial^b \X_\m $  (with $\ep_{\t\s} =-1$),  i.e. 
\be \la{tt}
\del_\t   \td \X_\m = - \frac{1}{Z^2} \partial_\s \X_\m \ , \ \ \ \ \ \
\del_\s   \td \X_\m = - \frac{1}{Z^2} \partial_\t\X_\m \ , \ \ \ \ \ \ \ \ \ \   \td Z =  \frac{1}{Z}= X_- \ . 
\ee
Thus $ (\td Z,\td\X_\m)$    satisfy the same set of equations \rf{bb1},\rf{bb2} 
corresponding   to the dual \ads{5}   metric  $ds^2 = {1 \over \td Z^2}   ( d \td Z^2  +  d \td \X_\m d\td \X^\m)$. 
Note that this T-duality is done along non-compact  target space directions that leads to subtleties when, 
as here,  %a1
the world sheet theory is defined on a cylinder (see Appendix C). 

The % transformation  in \rf{tt} may be written also      in terms  of the embedding coordinates $X_M$ in \rf{ep},\rf{p}  the 
T-duality relations in \rf{tt}  may be written also as 
\be 
 &&  \partial^a \tilde{\X}_\m  =  - \ep^{ab}\, j_{b,-\m}  \ ,  \la{ll}\\
  && 
  j_{a,  MN} \equiv X_M  \partial_a   X_N  - X_N  \partial_a   X_M \ , \ \ \ \ \ \   \del^a j_{a,MN}  =0 \ , \la{j}
 \ee
 where $ j_{a, MN}$ is the $SO(4,2)$ Noether current associated to the \ads{5} space symmetry which is  conserved on the equations of motion \rf{bb1}
 with the corresponding charge being \foot{Note that while in general 
  T-duality  interchanges  Noether charges with hidden conserved  charges,    the \ads{}  case is special 
 in that the  symmetries and thus the Noether charges of the original  and the dual backgrounds are isomorphic  (see \ci{rtw,bm,brtw}).} 
 \be 
 J_{MN}  = \int d \s \   j_{\t,  MN}   \la{jjj}  \ . \ee
 Thus the particular $SO(4,2)$ angular momentum components $J_{-\mu} $  
 get the   interpretation of the ``winding numbers''    in  the dual  $\td \X_\m$ directions, 
 \  $\int d \s\,  \del_\s \td  \X_\m=   J_{-\mu}  $.

   % just to  get familiar with the results. 
 
%While in general this is  just a conformal transformation that      gives an equivalent solution
%(though now with ``non-diagonal''   $SO(2,4)$ charges), 
%relating the parameter of the boost to a parameter of the solution  may lead to a new  solution in the limit
%as we shall see. 

%Under this transformation the  angular momentum components   entering 
%\rf{ll}    will get rescaled  accordingly .... so the winding will have interpretation of effective  light-like momentum 
%component...
%Relation of original and T-dual charges   in general   a la  \ci{rtw} ?  The simplest case os when the string moves in the subspace $X_4=0$
%since then $X_\pm=X_-1$.
 
  Let us  give  some 
 %In the following sections this is done with great detail but now we present a
  simple examples of  T-duality   applied to string solutions in  $AdS$. % without considering any special  limits.
  In general,  T-duality  formally relates a classical string solution to   another  solution  and
  thus  maps  the corresponding sets of conserved charges.
    For example, the  world  line   of  a  point-like  string or  massless geodesic in  \ads{}  is
      T-dual to  a  straight-line  Wilson loop surface  (in Minkowski world-sheet signature). 
      In more detail, the T-duality  counterparts     for    the three simple solutions in $\ads{2}\subset\ads{5}$ are (after T-duality it is convenient to interchange $\s$ and $\t$):
      %\foot{For       $\k$ all these solutions are the same.}
       
      % --  apparently not a widely known fact. 
  %et   us  start with solutions in $\ads{2}\subset\ads{5}$ and their T-duals. 
$\bullet$   massless    geodesic  in $AdS_2$  ($t= t(\t), \ \r=\r(\t)$, \ $\sinh \r=\k\t, \ \tan t = \k\t, \  \, \cos t = {1 \ov \sqrt{1 + \k^2 \t^2}} $)
\be 
Z=\T= { 1 \ov \k \t} \ , \ \ \ \ \ \ \ \ \ \ \ \ \ \ \ 
\td Z= \k \s \ , \ \ \ \ \ \    \td \T= \k \t  \ . \la{ge}
\ee
The   dual surface   reaches the boundary at $\s=0$ where it is a straight line along $\td \T \sim \t$. The reason we interchanges $\s$ and $\t$ after T-duality is precisely to have the new time coordinate $\td \T$ proportional to $\t$ (instead of $\s$). 
%a1 (assuming $\s$ is decompactified). 

 $\bullet$   massive  geodesic ($t= \k \t$, $\r=0$) or a  massless  BMN  geodesic in $AdS_2 \times S^1$
    (with $S^1$ angle $\varphi= \k \t$
 not involved in the T-duality transformation)
 \be 
 Z= {1 \ov \sin \k \t} \ , \ \ \ \ \   \T= \cot \k \t \ , \ \ \ \ \ \ \ \ \ \ \ \ 
 \td Z = \sin \k \s \ , \ \ \ \ 
 \td \T= \k \t  \la{g1} \ . 
 \ee
 The dual   surface  ends  at    the boundary at $ \s =0$   where it is  a straight line  along  $\td \T$
 and then again at  $\k \s= {\pi \ov 2}$. 
 %21 
 Notice that the dual surface has two boundaries corresponding to the intersection of the initial surface with the past and
 future Poincare Horizons.  
 %This is just a generalisation of straight-line surface to the presence of BMN   momentum $\k$.
 
 %A10 %21
 $\bullet$ wrapped string ($t=\kappa\t$, $\varphi=\kappa\s$  where $\varphi$ is an angle of a maximal circle of  $S^5$) \foot{The wrapped string solutions was considered, e.g.,  in \cite{RTT} (case $w=0$ in section 2.1 of that paper).}
 \be 
  Z= {1 \ov \sin \k \t} \ , \ \ \ \ \   \T= \cot \k \t \ , \ \ \ \ \ \ \ \ \ \ \ \ 
  \td Z = \sin \k \s \ , \ \ \ \ 
  \td \T= \k \t,\ \ \ \ \varphi=\kappa\t  \la{g11} \ . 
  \ee
  Here we again  interchanged $\t$ and $\s$ after  T-duality.

 $\bullet$   infinite straight static  string in $AdS_2$  ($t= \k \t, \ \r =\r (\s)$, $\tanh {\r \ov 2} = \tan { \k \s \ov 2}, $ $\cosh \r={ 1\ov \cos \k \s} $)\foot{This solution  may be viewed  as a special zero-spin limit \ci{FT} of a folded  spinning string in $AdS_3$.}
 \be 
 Z= {\cos \k\s \ov \sin \k \t} \ , \ \ \ \ \   \T= \cot \k \t \ , \ \ \ \ \ \ \ \ \ \ \ \ \ \ \ \ 
 \td Z ={ \sin \k \s \ov \cos \k \t}   \ , \ \ \ \ 
 \td \T=\tan  \k \t  \la{g2} \ . 
 \ee
 Here the original solution reaches  the boundary ($\r=\infty$)     at $\k\t= {\pi \ov 2} $   where 
 it ends on a  line  along $\T$. The dual solution   ends at the boundary for $\k\s={\pi \ov 2}n$  on a line 
 along $\td \T$.

 %25
It is clear   from  these examples   that to interpret the T-dual world surface as that of  an open string  stretched inside \ads{}
and ending  at the boundary at a point  (``quark'')   one needs to do two steps:

%S14
(i) Formally  {\it  ``decompactify''}   the $\s$ direction. 
%Since the T-duality relation  connects first derivatives of coordinates and thus  involves 
%integrals over $\s$ it may map  $\s$-periodic solutions  into  non-periodic 
%ones.  
Starting with a periodic solution for a small closed string (defined on a 2d cylinder) 
and applying the  T-duality (in a particular gauge, see below) 
 we  will get a $\s$-periodic  open-string world sheet, 
apart from the $\tilde \T$ direction  which gets  a term linear in $\s$. 
As discussed in Appendix C, the world-sheet theory   will  still  be periodic since $\td \T$ enters  only through its derivatives 
(assuming there are no  vertex operator insertions depending on $\td \T$).

%we will assume that its  T-dual  is  formally defined on a  half-plane (or on a strip). 
%In fact, in the T-dual wavy line solution discussed below  the   ``transverse''  coordinates   will  be  periodic  
%but  the  $\td \T$ direction   will    not.

(ii)    {\it  Interchange the $\t$ and $\s$   coordinates} 
(a familiar  operation in the  usual   flat-space closed--open string duality relation).\foot{Similar  transformation relating  oscillating and rigid spinning 
classes of string 
solutions in $AdS_5 \times S^5$  was 
 discussed  in a different  context  in \ci{hay}.}
%For example, in the massless   geodesic case \rf{ge} that will give  $\td Z= \k \s', \    \td \T= \k \t'$.
%We shall be doing a  similar   interchange  in small-string   examples    below. %  (omitting primes). 

 Let us briefly mention another example 
 %a1
% that will be the main topic of section 4
  -- the large spin limit of  folded spinning string  in $AdS_3$  ($ds^2 = - \cosh^2 \rho\ dt^2 + d \rho^2 + \sinh^2 \rho\ d\theta^2$)
 which is an   infinite  rotating  string stretching all the way to the boundary \ci{FT}:    $t=\theta=\k \t, \ \ \r= \k \s$, $\k \to \infty$.\foot{T-duality 
 applied to  finite-spin folded string  will be discussed in detail in section 4.}  
 This solution is   known to be related \ci{kt}, via a  Euclidean world-sheet continuation and an $SO(2,4)$ transformation, to the  null cusp 
 Wilson loop solution \ci{mk}.  The 
 T-duality provides a  different  relation to an open-string world sheet   ending at the boundary. 
 Indeed, 
 absorbing $\k$ into $\s$ and $\t$  and thus making their range infinite  we get  ($\X\equiv \X_1$) 
 \be  \la{lo}
&&   %{1\ov \cosh \s\,  \sin \t   - \sinh \s\,  \cos \t         } \ , \ \ \ \ \   
 Z =  (\cosh \s\,  \sin \t   - \sinh \s\,  \cos \t )^{-1}   , \ \  \T=  Z \cosh \s\, \cos \t , \ \   \X=  Z \sinh \s\, \sin \t \ ,      \\
 &&  \td Z ={     \cosh \s\,  \sin \t   - \sinh \s\,  \cos \t        }   \ , \ \ \ \td \T =\half \big( -\t + \s - \sin\t\,\cos \t + \sinh\s\,\cosh\s \big) \ , \cr 
 &&\td \X =\half \big( -\t -  \s   +  \sin\t\,\cos \t + \sinh\s\,\cosh\s \big) \ , \no \\
 &&  \td \T + \td \X =  -\t   + \sinh\s\,\cosh\s \ , \ \ \ \ \ \ 
 \td \T - \td \X =  \s  -   \sin\t\,\cos\t \ . 
  \la{loo}
 \ee
 The T-dual    surface ends at the boundary $\td Z =0$  
 when $\tan \t=\tanh\s$  where 
  \be
 \td \T + \td \X =  -  {\rm arctan } (\tanh \s)    + \ha \sinh 2 \s \ , \ \ \ \ \ \ 
  \td \T - \td \X =  \s  -   \ha \tanh 2 \s  \ .     % {\rm ArcTanh } (\tan \t)   - \ha \sin 2 \t
    \ee
    To  have  an open-string interpretation we are  again to replace $\s \leftrightarrow  \t$. 
   The  boundary trajectory  is a bended curve   passing through zero   in $(\td \T, \td \X)$ plane
   (at large $\t$ one  has $\td \T + \td \X  \sim e^{ 2 ( \td \T - \td \X )}$). 
   
%Xa = 1/2 ( - t - s   + 1/2 Sin[2 t]   + 1/2 Sinh[2 s])
%Ta = 1/2 ( - t + s   - 1/2 Sin[2 t]   + 1/2 Sinh[2 s])
%t = ArcTan[Tanh[s]]
%ParametricPlot[{Xa, Ta}, {s, -1, 1 }]
    %This does not look like null cusp ?   But we need euclidean world sheet to compare to null cusp and do not need T-duality ....
 %This i just some particular surface but plotting it is a good idea ...

 \
 
  Our  aim below will   be  to  apply  the  above  T-duality  transformation   to small  (``short'') strings in the bulk of  \ads{5}.
 To study short strings that effectively probe the near-flat limit of  \ads{5} 
  it is useful to consider a neighbourhood of 
 linear size $\epsilon\rightarrow0$ around the point $X_0=1$, $X_{M\neq0}=0$, namely 
 \be  X_0 \sim 1+ \epsilon^2\ , \ \ \ \ \ \ \ \ \ \ \ \ \ X_{M\neq0}\sim\epsilon \ . \la{nei}
 \ee
 An example  one may have in mind is  a small  string  of size $\ep$ oscillating near $\r=0$.
 In this case the components of  the Noether current  $j_{a, MN}$  in \rf{j}  scale as 
\beq
   j_{a,0M} \sim \partial_a X_M\sim\epsilon, \ \ \ \ \  j_{a,MN}\sim\epsilon^2, \ \ \ \ \ \ \ \ \ \mbox{for} \ \ M,N\neq 0\ . \la{214}
\eeq
In particular,  $j_{a,-0} \sim \epsilon$ whereas $j_{a, -i}\sim \epsilon^2$. 

In addition to this scaling limit, 
an important ingredient of our discussion   will be %At this stage we introduce an additional ingredient which is
a particular  $SO(4,2)$  transformation --  a large {\it boost}  in the 
 $(X_+,X_-)$   hyperbolic  plane  of the embedding 4-space in \rf{ep}
 \beq
X_-\rightarrow \frac{1}{\epsilon} X_-,\ \ \ \ \  \ \ \ X_+\rightarrow \epsilon X_+\ ,\ \ \ \  \ \  \ \ \   \epsilon\rightarrow 0 \ .  \la{bu}
\eeq
This boost makes a  small   string  near $\r=0$   move  fast towards the boundary, following  approximately a    massless 
geodesic. This   exposes  the  {\it near-BPS}  nature of a nearly point-like   small string state.  

 Assuming that  the $\ep$ parameters in \rf{nei} and in  \rf{bu} are the same, 
   the   resulting  scaling of the components of   $ j_{a,  MN}$   after the boost is 
\beq
  j_{a,-0}\sim 1,\ \ \ \  \  \ \ j_{a,-i},\,  j_{a,0i}\sim \epsilon,\ \ \ \  \ \ \ j_{a,+-},\, j_{a,ij},\, j_{a,+0}\sim\epsilon^2, \ \ \  \ \ \ \ j_{a,+i}\sim\epsilon^3\ . \la{216}
\eeq 
The components entering the T-duality relation in \rf{ll} are then % T-duality, the relevant currents are
\beq
 j_{a,-0} \simeq \del_a X_- \sim   1 ,\ \ \ \ \  \ \ \ \ \ \ j_{a,-i} = X_-\del_a X_i - X_i \del_a X_- \sim  \epsilon \ , \la{217}
\eeq
implying that  for $J_{MN}$   in \rf{jj}  one has  $J_{-0}\sim 1$ and $J_{-i}\sim \epsilon$.
This  means, in the T-dual interpretation, that the dual string is extended
along $\tilde{\X}_0=\tilde{\T}$ with fluctuations of order $\epsilon $ in the directions $\tilde{\X}_{i=1,2,3}$. 
Since   $X_-\sim 1$   the dual  string  world sheet  extends also along $\tilde{Z} = 1/Z = X_-$.

 It is useful  to record  the scaling of the  conserved charges    representing the    energy and the spin of a  short string  we started  
 with (we omit  the   factor of string tension, i.e. $E= {\sql \ov 2 \pi} \bar E$, etc.)\foot{Note also that $  \bar  \S_{4i} \equiv  J_{-i} - J_{+i}  = \cO(\epsilon)  $ 
  may be  interpreted,  from the point of view of dual 
   conformal  field theory  on $\mathbb{R}^{3,1}$,  as a
  linear combination of the momentum  generator and  the generator   of special conformal transformations.} 
\beqa
 \bar  E &\equiv & J_{-0} + J_{+0} = \int\! d\s\, \del_\t X_- + \cO(\epsilon^2)\ , \la{218}  \\
  \bar \S_{ij} &\equiv & J_{ij} = \cO(\epsilon^2) \  .   \la{219} 
\eeqa

 To summarise, 
 %In conclusion, 
 (i) expanding around the point $X_0=1$, $X_{M\neq0}=0$ in \ads{5}   and (ii) performing  a large  boost in the $X_\pm$ plane 
 with the same   parameter $\ep$ as the expansion scale
   produces a   string   configuration   which has the   T-dual 
 interpretation as  a  world sheet extended along $\tilde{\T}$, $\tilde{Z}$ with small fluctuations in the  
 spatial $\tilde{\X}_{i=1,2,3}$   directions. 
 We shall   study this  relation    in detail in the following   section.

\renewcommand{\theequation}{3.\arabic{equation}}
\setcounter{equation}{0}

 \section{T-duality relation between   short   closed strings  and  long wavy open strings }
 
 Let us elaborate on the   small   string expansion  discussed  above. Starting with global coordinates in \rf{nk}  we may set 
 % Consider the limit of a small string  inside AdS. 
% A way to do it in global coordinates is to rescale 
 \beq
  \rho = \epsilon \bar{\rho} ,\ \ \ \ \ \  \ \ t=\epsilon  y_0 \ , \la{3.1} 
  \eeq
% In the metric
so that   the metric becomes  nearly flat
  \beqa\la{51} 
  ds^2 &=& - \cosh^2\rho dt^2 + d\rho^2 + \sinh^2\rho d\Omega_{[3]}^2 \cr
     %  &=& -\epsilon^2  \cosh^2(\epsilon\bar{\rho}) d\bar{t}^2 + \epsilon^2 d\bar{\rho}^2 + \sinh^2(\epsilon \bar{\rho}) d\Omega_{[3]}^2 \cr
     %  &\simeq & \epsilon^2 \left( - d\bar{t}^2 + d\bar{\rho}^2 +  \bar{\rho}^2 d\Omega_{[3]}^2 \right)\cr
       &\simeq&  \epsilon^2 \left( - d y_0^2 + dy_r dy_r \right)  \ , \ \ \ \ \ \ \ \  y_r = \bar{\rho} \hat{n}_r\ , 
  \eeqa
  where % we defined $y_a = \bar{\rho} \hat{n}_r$, $r=1, \cdots, 4$  with  
  $\hat{n}_r$  is  the unit vector in \rf{nnk}. 
% not surprisingly giving flat space in the leading approximation. It is convenient to parameterize the sphere $S^3$ by a unit vector $\hat{n}\in\mathbb{R}^4$
% and define $y_a = \bar{\rho} \hat{n}_a$, $a=1\cdots 4$.

 \def \lar {\leftrightarrow} 
 
The corresponding small string equations   will  be  the same   as in flat space.   Let us fix  the conformal gauge  and 
%Let us denote $y_0=\bar{t}$.   We shall use conformal gauge   where T-duality transformation looks simplest. 
 %Let us 
%22
%a1 a 
further choose  a particular  conformal  parametrisation    (light-cone gauge)   where 
 \beq
   y_- \equiv y_0 - y_4 = \tau  \ . \la{52} 
 \eeq
 Then the solution for the transverse spatial directions  $y_i$   
 is  the usual   flat-space one  % (may be can be done at the quantum level ?)
 \be
   y_i (\sigma,\tau ) = \ry_i(\s,\t) \equiv \ry_i^{+}( \sigma+\tau) + \ry_i^{-}(\sigma-\tau) \ , \  \ \ \ \ \ \ \ 
  \ \  \ i=1,2, 3 \la{53} 
  \ee
%a1
with an initial value at $\t=0$ given by 
\be
   y_i(\sigma,0  ) = \ry_i (\s) = \ry_i^{+}( \sigma) + \ry_i^{-}(\sigma) \ . \la{553}
 \ee
% If this is a closed-string solution it should  also satisfy the standard flat-space   Virasoro  conditions, 
 %22 {\bf fix coeffs } 
% \be \la{vir}
% \del_+ y_+ = ( \del_+  \ry^+)^2 \ , \ \ \ \ \ \ \ \ \  
%  \del_- y_+ = ( \del_-  \ry^-)^2  \ , \ee
 % implying, e.g., standard level matching condition $\int ^{2\pi}_0 d \s  [ ( \del_+  \ry^+)^2  - ( \del_-  \ry^-)^2 ]=0 $.
   As we shall explain  below, 
% The statement is that
 this leading-order small-string configuration  is  T-dual %, to the  leading order, 
 to   an open-string   solution  ending  at the boundary along 
 the following trajectory  (after $\t\lar \s$ interchange) 
 %\foot{Note that    we used $y_4$  direction  to fix  light-cone gauge,  $y_i$  represents 
 %one dimension less, i.e.  corresponds to transverse directions  at the boundary.} 
 %represented by  % Wilson loop
 \be
  \X_0 =\t\ ,\ \ \ \ \ \    \ \ \ \ \ \ \ \   \X_i = \rx_i(\t) = \int^\t   d\t'    \, \ry_i(\t')    \ .       \la{54} 
 \ee
 It thus  represents  a small deformation  of the straight-line Wilson loop $\X_0=\t$.

 %22
 Let   us recall how the   small-string limit looks in the 
  embedding coordinates \rf{nei}  -- as an   expansion around the point $X_0=1$, $X_\pm=X_i=0$:
 \be
X_0 \sim 1 +\epsilon^2 , \ \ \ \ \ \   X_\pm,X_i \sim \epsilon ,  \ \ \ \ \  \ep \to 0 \ . \la{33} 
\ee
The equation of motion for $X_0$ in \rf{bb1}  implies that  $\Lambda \sim \epsilon^2$
and  the constraint $X_+ X_- -  X_\mu X^\mu=1$  determines the 
$\epsilon^2$ term in $X_0$.  To  the lowest order in $\ep$   in  each of the equations in \rf{bb1},\rf{bb2} 
 we find
\be
&&  \pa_a \pa^a X^i =0, \ \ \ \ \ \ \  \pa_a   \pa^a X_\pm = 0, \la{bb3} \\
&&
   \pa_\s X_+ \pa_\s X_- +\pa_\t X_+ \pa_\t X_- - \pa_\s X_i \pa_\s X_i - \pa_\t X_i \pa_\t X_i =0\ , \no \\
  &&
  \pa_\s X_+ \pa_\t X_- +\pa_\t X_+ \pa_\s X_- - 2  \pa_\s X_i \pa_\t X_i =0 \ . \la{bb41} 
  \ee
%These are, as expected, the appropriate equations for a string in flat space.
These are, once again, the equations for a string in flat space.  Fixing 
 the residual conformal reparametrization symmetry  by the  light-cone gauge\foot{Let us 
 stress   that this relation refers only to the leading term in the $\epsilon$ expansion of $X_-$.}
\beq
X_-=\ep \tau \ ,  \la{lc} 
\eeq
  the constraints  become   %are solved by taking
\beqa
 \ep\, \pa_\t X_+ =  \pa_\s X_i \pa_\s X_i + \pa_\t X_i \pa_\t X_i \ , \ \ \ \  \ \ \ \ \ \ 
 \ep\,  \pa_\s X_+  = 2 \pa_\s X_i \pa_\t X_i \  \la{bb5} 
 \eeqa 
 and are readily solved as usual. 
%MK: 
 The period $\Sigma$ of $\s$, i.e.  $\s\equiv \s + \Sigma$ can be found by computing the energy from eq.(\ref{218}).
\beq
 \bar{E} = \bar{P}_- = \int d\s\  \pa_\t X_- =  \epsilon \Sigma \ . %  \ \ \ \Rightarrow 
 %\ \ \ \Sigma %= \frac{\bar{E}}{\ep}=\frac{2\pi}{\sqrt{\lambda}} \frac{E}{\ep}
\label{Speriod}
\eeq 
 Notice that the period is finite since $\bar{E}\sim\ep$ for a string of size $~\ep$. 
 %{\bf This is before the boost, may be it is better to define $\bar{E}$ after the boost? }
 After the boost \rf{bu}  that   removes the factor of $\ep$   from $X_-$  (and thus from  $E$) we   will   have 
 \be 
 \Sigma = {\bar{E}}=\frac{2\pi}{\sqrt{\lambda}} {E}  \ . \la{spe}
  \ee
 %which are compatible in view of the equations of motion. 
%The only extra information  about  $X_i$ is their 
The  periodicity of $X_+$  in $\s$    implies 
\be
&& 0 =\ep\,  \int d\s\ \pa_\s X_+ = 2 \int d\s \, \pa_\s X_i \pa_\t X_i = 2 \ep^2  \int d\s\,  \left[ (\pa_\t \ry^+_i )^2 - (\pa_\t \ry^-_i )^2 \right] \ , \la{bb6} \\
&&  X_i =\ep  y_i =\ep \ry_i   \   , \ \ \ \ \ \ \ \ \ \  \ry_i (\sigma,\tau ) = \ry_i^{+}( \sigma+\tau) + \ry_i^{-}(\sigma-\tau)  \ , \la{34}
\ee
where, as in \rf{53}, the functions $\ry_i^\pm$ are the left and right moving parts of the harmonic functions $\ry_i$ 
representing the solution for the transverse coordinates.
   This is the 
usual flat-space level matching condition  which forbids, for periodic strings, 
 the existence of states with  purely left or right moving modes. 
%This will be useful later when describing the T-dual Wilson loop.\foot{Notice
 %also that as part of the derivation we later on do a boost in 
%direction $X_\pm$. 
%Since the product $X_+ X_-$ remains invariant the calculation is still valid. }

 To   summarize,    the small-string limit     in  the embedding coordinates reads, to the  leading order, as  (cf. \rf{nei},\rf{3.1},\rf{52},\rf{33},\rf{lc})
 \be
   X_0 = 1\ ,   \ \ \ \ \ \   
   X_i = \epsilon \bar{\rho} \, \hat{n}_i =\epsilon  y_i, \ \ \ \ \ \ \ \ \ % i=1\cdots 3 \ , \ \ \ \ \ 
   X_\pm  = \epsilon ( y_0\pm   \bar{\rho}\, \hat{n}_4) = \epsilon ( y_0 \pm   y_4) \la{57}\ . 
   %X_+ &=& \epsilon ( \tilde{t} +  \tilde{\rho} \hat{n}_4) = \epsilon ( \tilde{t} +  y_4)
  \ee
Let us  now  write the corresponding solution   in   the  Poincare coordinates  \rf{p1}--\rf{pok}, i.e. 
 \be
  Z = \frac{1}{X_-} \ , \ \ \ \ \ \ \ \ \ \ \ \ 
  \Y_\mu =  \frac{X_\mu}{X_-}\ , \ \  \  \ \    \   \mu=0, \cdots,  3\ .  \la{56} 
  \ee
 We find 
\be
   Z =  \frac{1}{\ep\,(y_0 -  y_4)} \ , \ \ \ \ \ \ \ \ 
 %  \Y_0 &=& Z =\frac{1}{\epsilon} \frac{1}{y_0 -  y_4}\\
 \Y_0\equiv \T = \frac{1}{\ep\, (y_0 -  y_4)}    \ , \ \ \ \ \ \ \ \ \ \  \Y_i =   \frac{ y_i}{y_0 -  y_4}\ .   \la{58}\ \  % \mu=1\cdots 3
\ee
Next, let us  perform the boost \rf{bu} 
%AT as in \rf{bus},\rf{re},     the simple but 
  %an important rescaling (boost)
   by multiplying
 all the  Poincare  coordinates by $\epsilon$ (which is an obvious symmetry of the metric).  This gives 
\be
   Z =\T=  \frac{1}{y_-} \ , \ \ \ \ \ \ \   \ \ \ \ \   %\\
%   \Y_0 &=& Z = \frac{1}{y_-}\\
   \Y_i= \frac{\epsilon\, y_i}{y_-}\  , \ \ \ \ \ \ \ \    y_-\equiv y_0-y_4 \ .   \la{59} % \ \ \mu=1\cdots 3
\ee
%AT
Fixing also the light-cone gauge parametrization as in \rf{52},\rf{lc}  we conclude   that 
this   configuration   represents, to  the leading  order  in $\ep\ll 1$, the    boosted small-string 
 solution  in  the Poincare  patch. % the  conformal-gauge    solution. 
% (as in \rf{ne}  where $k=\ep$). 
 Indeed, it  is easy to  check  directly    that the   equations for $Z$   and $\T$ are solved to leading order in $\ep$ %small $y_i$ expansion, 
 while  the equation for $\Y_i$, i.e.     $\del_\t ( Z^{-2} \del_\t  \Y_i) - \del_\s ( Z^{-2} \del_\s  \Y_i) =0$,  is 
 also     satisfied provided that  $y_i=\ry_i$   as in \rf{53}  where $\ry_i$  is harmonic function.

%22
In what follows we shall   absorb the   $\ep$  factor in \rf{59} into $y_i$ 
  assuming that  we always expand to leading order in small $y_i$. 
 Then for  the above  flat space solution \rf{52},\rf{53}  %  in the light-cone gauge $y_- =   \tau$ we thus get 
we get from \rf{59}  the following leading-order solution in the Poincare coordinates
 \beqa
    Z = \T =   \frac{1}{\tau} \ , \ \ \ \ \ \ \ \ \ \ \ 
   % \Y_0 &=& Z = \frac{1}{\tau}\\
    \Y_i=  \frac{ y_i(\sigma,\t)}{\tau}\  , \ \ \ \ \ \ \ \ \     y_i= \ry_i(\s,\t) \   .   \la{60} \ % \ \mu=1\cdots 3
 \ee

 \subsection{T-duality  transformation}
 
 Let us now  apply   the T-duality  transformation    \rf{tt} to the  small-string    solution \rf{60}. We find 
  %which is quite simple
  \beqa
   \tilde{Z} &=& \frac{1}{Z} = \tau \ , \la{16}  \\
   \partial_\sigma \tilde{\T} &=& - \frac{1}{Z^2} \partial_\tau \T = 1\ , \la{116}  \\
   \partial_\sigma \tilde{\Y}_i &=& - \frac{1}{Z^2} \partial_\tau \Y_i = -\tau^2 \partial_\tau\frac{y_i}{\tau}=  y_i -\tau\partial_\tau y_i 
    \ . \la{jj}
  \la{61}  \ee
   We thus get the following  leading-order solution in the  dual  Poincare  patch 
   %to leading order in $\ep \ll 1$
     (removing   the tildes on the dual coordinates)
 \be
  Z(\sigma,\tau) &=& \tau \ ,  \ \ \ \ \ \ \ \ \ \ \ \ \ \ \ \ \ 
  \T(\sigma,\tau) =  \sigma  \ ,  \la{66}  \\
  \Y_ i (\sigma,\tau) &=& \int^\sigma  d \s'\, \Big[ y_i(\s', \tau) - \tau \partial_\tau %\int^\sigma  d \s'\,
   y_i(\s', \tau) \Big] \label{ysol}  \ , 
  \la{62}  \ee
  where $y_i=\ry_i $ is the   harmonic function in \rf{53}.
  This  surface ends  at the boundary $Z=0$  at   $\tau=0$  with  the  shape  of the boundary curve being 
  %The shape of the Wilson loops is 
  \be
   \T =  \sigma \ , \ \ \ \ \ \ \ \ \ \ \ 
   \Y_i(\sigma) =  \int^\sigma d \s' \,  y_i(\s', \tau=0) =    \int^\sigma d \s' \,   \big[  \ry^+ _i(\s') +  \ry^- _i(\s') \big]   \ . 
 \la{63}   \ee
 %where we remind the reader that $y_i$ is a solution in flat space in light-cone gauge. 
 The third normal  derivative  is  given by
 \beq
  \partial^3_\tau\, \Y_i(\sigma,0) = -2    \int^\sigma d \s' \,  \partial^3_\tau y_i(\s', \tau=0) 
  = -2   \int^\sigma d \s' \,  \partial^3_{\s'}  \big[  \ry^+ _i(\s') - \ry^- _i(\s') \big]  \ . 
  \label{63c}
 \eeq
 Equivalently, eqs. (\ref{63}),(\ref{63c}) as boundary conditions  completely determine (\ref{66}),(\ref{62}) as the unique solution.
 
 Let us now compare   this to the  small-wave open-string  solution of \ci{mik}   we reviewed   in \rf{22},\rf{23}.
 To this end we shall relax the condition of periodicity   in $\s$ that was implicitly assumed for small strings in the middle of \ads{5} 
 and then  {\it interchange}  $\tau$ and $\sigma$,  thus  writing    \rf{66},\rf{62}    as 
  \be 
 Z&=& \sigma \ , \ \ \ \    \ \ \ \ \ \   \T=\tau \ , \ \ \ \  \ \ \ \ \   \Y_ i  = x_i (\t,\s) \ ,     \la{76}\\
 x_ i (\tau,\s)& =& \epsilon \int^\t  d \t'\, \Big(   \ry_i^{+}( \tau+ \s) + \ry_i^{-}(\t - \sigma)    
           - \s   \big[  \dot \ry_i^{+}( \tau+ \s) - \dot  \ry_i^{-}(\t - \sigma)  \big]           \Big)
   \label{567}  \  . 
 \ee
This  is, indeed,  equivalent to \rf{111},\rf{22} if we set (cf. \rf{54}) 
\be 
  \ry^\pm _i (\tau) = \dot \rx^\pm_i (\tau)  \ . \la{68} \ee 
%AT
 %Surprisingly,
  We observe    that  the original small-string profile is mapped into the  derivative ({\it velocity})
 of the open-string  end point   at the boundary,  and thus  the  expansion in {\it  small string size} %$\ep$ 
 corresponds  to   the {\it small velocity}   expansion  for  the wavy line.\foot{Since we started from a flat-space solution  corresponding to  the  small size ($\epsilon \rightarrow 0$) limit 
 we should not get   an exact solution  after the T-duality.
 % approximation and discarding higher orders in $\epsilon$.
 Indeed, the dual solution 
 %What we are going to see is that the solution
  is a solution to the linearized equations of motion in $\ads{5}$ \ci{mik}.   The leading-order solution is expressed in terms  of 
    a superposition of  left and right moving waves;
       corrections to it 
        can be computed by going to higher order in the $\epsilon\rightarrow 0$ expansion.}  
Notice that since we interchanged $\tau$ and $\s$  the world surface  here is periodic in $\tau$ with period  $\bar E$ 
equal to the small    string energy  (see \rf{spe}).

% This
% Now we should 
% checks that  we indeed found %somehow corresponds to
 % a solution for a string in $\ads{5}$ as  it should   be  under T-duality.
  %not   get  an exact solution 
     This proves that an arbitrary
  flat space solution maps, under  our  ``T-duality  plus boost''
  transformation into a linearized solution for an open  string ending on the boundary  along a wavy small-velocity trajectory. 
 % Note that the  $\Y_i$ (or $y_i$)  having extra factor of $\ep$   are  thus  considered to be small:
 Since  a deviation   from the straight line is small, 
 this may be   interpreted as a  near BPS  configuration. 
 The same  applies to the original  small string: as  was already mentioned above, the large boost
 makes its  world surface   similar to that of a null geodesic in \ads{5}.

%  since we expand around the straight line  Wilson loop  surface.\foot{Thanks to the boost the original small closed 
  %string became near BPS.}  
  
 Below we   shall  compute    the   value of the action (area) of this dual open   string solution and the corresponding energy and spin. 
 We shall find that  the string  energy is given in terms of its profile by \rf{er},\rf{iy}. 
  In a sense,  the    T-duality  relation helps to ``demystify'' the relation between  the string 
  energy  (depending usually on  first   derivatives of  the coordinates) and the Larmor-Lienard expression    \rf{er} 
  of \ci{mik}   that involves  the second derivative (acceleration) of the boundary   trajectory. 
 
 It is reasonable to expect that the  closed-open string relation    we discussed   above 
  extends  beyond the classical approximation, 
 i.e.  applies to all orders in the  large  string tension expansion (expansion in $1\ov \sqrt \lambda$). 
 This is suggested by the  fact that the world-sheet theory is the same,
  the difference is in the interpretation of  its  fields  (which are related by T-duality). 
  At the same time, the  relation   as described above is  still 
   valid only  in the small-fluctuation (i.e. small velocity)  approximation. 
    To extend it beyond the leading order   in 
  % It should be possible to extend this  relation beyond 
 this approximation remains  an interesting open problem. 
 %requires careful     analysis of  how the anergy is comhow the energy is computed on the Wilson loop side. 

 \subsection{Area of the open string surface}
 
 Let us   consider a Wilson loop which is a small deformation of the straight line and 
  compute  the  area of the corresponding minimal surface   determining its value at strong coupling.
    % As was 
  As was already mentioned above, since we assume  the Minkowski  signature on  the world sheet,  
  we need to specify not only the  boundary shape but also the third normal  derivative  of
  the  transverse  coordinates of the surface. 
  We shall also assume that the  integrals of partial derivatives along the Wilson loop vanish
%  \ie\ either 
  (which is the case  if asymptotically the Wilson loop is a straight line or if we consider  only periodic perturbations). 
  The expression   for the  area of the surface  \rf{76}  (where $x_i$   may be viewed as   a generic small-fluctuation  solution) 
    in the  leading  approximation  is given by\foot{We  ignore the trivial 
  divergent $x_i$-independent term.}
\beq
   S = {\sql \ov 2\pi} \bS \ , 
   \ \ \ \ \ \ \ \      \
   \bS=  \int  \frac{d\tau d\s}{2\s^2} \left[ (\partial_\tau x_i)^2 - (\partial_\s x_i)^2 \right] \la{b1}\ . 
\eeq
Using the equations of motion and dropping total derivatives  in  $\tau$  we get
\beq
 \bS =\left. \int  \frac{d\t}{2\s^2}\,  (x_i \partial_\s  x_i ) \right|_{\s=\varepsilon}^{\infty} 
   = - \left. \frac{1}{2\varepsilon^2} \int d\tau\,   ( x_i \partial_\s  x_i ) \right|_{\s=\varepsilon} \ , \la{1b} 
\eeq 
 where we introduced a near-boundary cut-off $Z=\s=\varepsilon\to 0$ \foot{$\varepsilon$ here should 
 not to be confused with the $\epsilon$ parameter  of the expansion we used in section3.} and 
  dropped the $\sigma\rightarrow \infty$  term  under 
   the assumption that $x_i$ and $\partial_\s x_i$ are finite everywhere.  
   Expanding in small $\s$  we find 
   % near boundary expansion:
  \be 
( x_i \partial_\s  x_i )_{\s=\varepsilon}&=&
( x_i (\t, 0)   + \ha  \epp^2  \del^2_\s x_i  (\t, 0) + ...)   ( \epp \del^2_\s x_i  (\t, 0)   +   \ha \epp^2   \del^3_\s x_i  (\t, 0)  + ...) \no\\
&=& \epp x_i (\t, 0)   \del^2_\s x_i  (\t, 0)    +  \ha \epp^2   x_i (\t, 0) \del^3_\s x_i  (\t, 0)  + ... \la{b2}\ . 
 \ee
 Using that   for the solution of equations of motion \rf{22}  we have 
  $x_i =\rx_i - \s \del_\s \rx_i  +   O(\rx^2)$   where $\rx_i$ is a harmonic function  
 we  obtain  (to  leading  order)\foot{Note that 
 %$ x_i \partial_\s  x_i  = (\rx_i - \s \del_\s \rx_i  ) ( - \s \del^2_\s  \rx_i) $.
 %  we have 
 here the first normal derivative is zero, the second one is  dependent on boundary curve and third one  is  finally independent. 
% Thus the second normal derivative is
  }
 \be 
 &&  x_i \Big|_{\s=0}   =    \rx_i\Big|_{\s=0} = \rx_i(\t)  \ , \ \  \ \ 
    \del_\s x_i \Big|_{\s=0} =0 \ , \ \   \ \    \del^2_\s x_i \Big|_{\s=0} =  - \del_\s^2 \rx_i \Big|_{\s=0}= -   \del_\t^2 \rx_i (\t) \ , %= - \del_\t^2  x_i \Big|_{\s=0} 
      \no \\
    && \ \ \ 
  \del^3_\s x_i \Big|_{\s=0} =  -  2  \del_\s^3 \rx_i \Big|_{\s=0},  \ \   \ \ \ \ \  \pa^k_\t x_i \Big|_{\s=0}=\pa^k_\t \rx_i(\t) \ . \la{bb4}
 \ee 
 Thus   $ \del^2_\s x_i  (\t, 0)  = - \del^2_\t x_i  (\t, 0)  $    so that   after integrating by parts
\beq
  \bS = - \frac{1}{2\varepsilon} \int d\tau\,  [\partial_\tau \rx_i(\tau) ]^2 - {1 \ov 4}  \int d\tau\,  \rx_i  (\tau)  ( \partial_\s^3 x_i) (\tau,0)  \ . \la{b4} 
\eeq
To compute the area  we thus need both boundary conditions, namely, the boundary shape and the third
normal  derivative. %These  are two independent functions    of $\t$; 
Equivalently, one may choose as the  independent  boundary data 
 the left and right parts $\rx_i^\pm(\tau)$   of the harmonic function $\rx_i(\t, \s)$.
 %This is  the reflection of the general fact that in Minkowski    signature we need  {\it two independent boundary data  (functions of $\t$)}
% to specify a solution  for world surface. 

  The first divergent term in \rf{b4} gives a correction to  the length of the Wilson loop. 
  The finite part of the area comes from the second term
\beq
  \bS_{\rm fin}  =  - \fo  \int d\tau\,  x_i (\tau,0)  (\partial_\s^3 x_i) (\tau,0) \ . \la{b5} 
\eeq
 This can be written  in terms of the   parts $\rx^\pm_i=\rx_i^\pm(\tau)$ of the 
 harmonic function $\rx_i(\t, \s) = \rx^+_i(\t +  \s) + \rx^-_i(\t- \s) $  as (see \rf{23}) 
\beq
  \bS_{\rm fin}   = \ha   \int d\tau\,   (\rx_i^+ + \rx_i^-  )   (  \del^3_\t \rx_i^+  -  \del^3_\t   \rx_i^-  ) 
  = \ha \int d\t \ v_i(a_i^--a_i^+)
%=\ha   \int d\tau\,     ( \rx_i^- \del^3_\t   \rx_i^+  - \rx_i^+ \del^3_\t   \rx_i^-)
 \ ,  \la{b6}
\eeq
% I changed sign of last term.
where   we  integrated  by parts\foot{We dropped boundary terms assuming that   at $\t=\pm \infty$  the 
contour is straight   line and the 3rd derivative vanishes, i.e. $ \rx^\pm_i (\tau \to \pm \infty) =0$.} 
 and introduced the  obvious  definitions
\beq
 v_i = \pa_\t \rx_i(\t,0) ,\ \ \ \ \ \    \ \ \ a_i^\pm=\pa^2_\t \rx^\pm_i(\t)\  . \la{df}
 \eeq
%The area    thus vanishes if one has only  the  left or only the right wave   propagating from the boundary to the bulk. 

 %%%%%%%%%%%%%%%%%%%%%%%%%%%%%%
 
 \subsection{Energy of the  wavy line open string}
 
 Let us   now consider the energy of the open 
  string that ends on   the boundary at a point which is  fluctuating around the straight Wilson 
 line $\T=\tau$  with the transverse coordinates $\rx_i=\rx_i(\t)$ changing slowly. 
  This end point may  be referred to as a ``quark''. The energy is time dependent because the work 
   needed to keep the quark on  its trajectory is radiated as waves into the attached  string.
 There may be also  waves that  arrive  from the string   and  are  absorbed by the quark. 
 
 The  case of pure emission  by the quark  was studied by Mikhailov
 \ci{mik}  who obtained a very interesting result for the energy of the string at a time $\T$  (see also \ci{che})   
\beqa
 E(\T) &=& \frac{\sqrt{\lambda}}{2\pi} \bar{E}(\T) \ , \la{e1}  \\
 \bar{E}(\T) &=&  \frac{1}{\varepsilon} \frac{1}{\sqrt{1-v^2}}- 
      \frac{v\cdot  a}{(1-v^2)^2}
   +   \int_{-\infty}^\T\!\! d\t\ \frac{(1 - v^2)\, a^2+ (v\cdot a)^2}{(1-v^2)^3}\ , 
   \label{E1}
\eeqa
where $v_i = \pa_\t \rx_i(\t)$, $a_i=\pa_\t^2 \rx_i(\t)$ and in the first two terms $v_i= v_i(\T), \ a_i=a_i(\T)$. 
 In \ci{mik} only the last term in \rf{E1} was derived explicitly  and  it  was integrated over the whole trajectory. 
 %Here it is  convenient to include  
 The first two terms come  from the regularization 
near the boundary, i.e. at 
 $Z=\s=\varepsilon\to 0  $. The last term is the accumulated  energy that the quark radiates into the string and  it   should  therefore  be 
always increasing (the integrand is indeed positive).  
In the leading small velocity approximation eq.(\ref{E1}) reduces to
\beq
 \bar{E}(\T) =  \frac{1}{\varepsilon} (1+\ha v^2) - v\cdot a+   \int_{-\infty}^\T\!\!\! d\t\  a^2 (\t)  \ . \la{e2}
\eeq
 This  expression  can be conveniently  encoded   in terms of the
   value of the first derivative   of the function $\bar E(\T)$:  % or, equivalently, over $\t$ 
 \beq
  \frac{\partial \bar{E}}{\partial  \T } =  \frac{1}{\varepsilon} v\cdot a -\pa_\T(v\cdot a) +  a^2 \ . \la{e3}
  \label{E25}
\eeq
Let us   now  derive this expression by  starting directly   with the  small-wave solution in \rf{22}.
For generality, we will   keep both $\rx^+$ and $\rx^-$ modes, i.e.  will  not 
make an a priori   assumption that there are only outgoing waves. 
In the leading-order  approximation   the  string energy  is 
(here $E= \frac{\sqrt{\lambda}}{2\pi} \bar{E}$  is   conjugate to $\X_0=\T$  equal 
in the case of   $\T=\t$  in \rf{76} to the   2d  open string energy) 
\beq
 \bar{E}(\tau)  = \bar E^{\rm op} = \int_{\varepsilon}^\infty \frac{d\s}{\s^2} \Big[ 1+\ha(\partial_\s x_i)^2+\ha(\partial_\t x_i)^2\Big] \ . \la{e4}
 \eeq
This  implies that 
 \beq
 \partial_\t \bar{E} =  \int_{\varepsilon}^\infty \frac{d\s}{\s^2} \big(\pa_\s  x_i\pa^2_{\s\t} x_i+\pa_\t  x_i\pa_\t^2 x_i\big)   \ . \la{e5}
 \eeq
 Using the equations of motion 
\beq
  \pa_\t^2  x_i - \pa_\s^2  x_i+ \frac{2}{\s} \pa_\s  x_i =0\  , \la{e6} 
  \eeq
it follows that the integrand is a total derivative (as expected from the fact that it comes from a conserved current). 
Therefore,  the energy change is due
to an energy flux coming from the boundary and is given by
\beq
 \pa_\t \bar{E} = -\left.  \frac{\pa_\s x_i \pa_\t x_i}{\s^2}  \right|_{\s=\varepsilon} \ . \la{e7} 
 \eeq
Using the expansion  following from  \rf{bb2}
\beq  
 \partial_\s  x_i(\t,\varepsilon) = - \varepsilon  \partial_\tau^2  x_i (\tau,0) + \ha \varepsilon^2 \partial_\s^3 x_i(\tau,0)\ , \la{e8}
 \eeq
we find 
\beqa
  \pa_\t \bar{E} = \big(\frac{1}{\varepsilon} \pa_\t  x_i \pa_\t^2  x_i - \ha \pa_\t  x_i \pa_\s^3  x_i \big) (\tau,0)  
          =  \frac{1}{\varepsilon} v_i a_i - \ha   v_i  \pa_\s^3  x_i  (\tau,0) \ , \la{e9}
\eeqa
where once again $  v_i= \del_\t \rx_i(\t) = \del_\t x_i(\t,0)$, etc. 
%  everything is evaluated at the boundary $\s=0$ and depends on $\t$. 

Comparing  \rf{e9} to \rf{E25}  we see that the 
 divergent term is reproduced,   but  the finite term depends on the third derivative. 
 Therefore,  the balance of the energy absorbed and emitted by the quark depends, as expected, 
 on both boundary conditions.  
 Writing the solution in terms of the left and right moving waves we have  as in \rf{22},\rf{23}  
\beqa
  x_i(\t,0) = \rx^+_{i}(\t)+ \rx^-_{i}(\t)  \ , \ \ \ \ \ \ \ \ \    
  \ps^3 x_i(\t,0) = -2[\pa_\t^3\rx^+_{i}(\t)-\pa_\t^3 \rx^-_{i}(\t)]  \la{e10} \ , 
\eeqa
and  thus 
\beq
 \pa_\t \bar{E} =  \frac{1}{\varepsilon} v_i a_i +   v_i(\dot a^+_i-\dot a^-_i)   =
     \frac{1}{\varepsilon} v\cdot a +  \pa_\t\Big[v\cdot (a^+ -a^-)\Big]
              +  a^-{}^2  - a^+{}^2  \ , \la{e11} 
\eeq
where   we used   the   definitions
\beq
v_i = \pa_\t\rx_i=\pa_\t\rx^+_i + \pa_\t\rx^-_i\ ,  \ \  \  \ \ \  a_i= a^+_i + a^-_ i \  , \ \  \ \ \ \ \ a^\pm _i= \partial_\t^2 \rx^\pm _{i}  \ .  \la{e12}
 %\ \ \ a^+_i=\pa_\t^2 \rx^+_{i}
\eeq
In the case of pure emission   described by the  special solution with 
$\rx_i^+=0$  (i.e. with $a=a^-$)  we conclude that  the  full  expression in  eq.(\ref{E25}) is  reproduced.

In the general case when 
 there are  both  the left and the  right moving waves, the ``interference''
  term in \rf{e11}\foot{Note that this term is  the same 
  as   the integrand in the area in \rf{b6}.}
  is a total derivative
  and the logical  conclusion  follows:  the energy of the string increases when the quark radiates  the energy  into the string 
 (the term $(a^-)^2>0$)  and decreases when the quark absorbs the energy  from the   string  (the term $- (a^+)^2<0$).
 
 The result can be expressed 
 in terms of the variables of the T-dual  short string (see \rf{66}--\rf{68}). 
 The  small string  solution is essentially 
 the flat space solution expressed in terms of  the functions $\ry^+(\t+\s)$ and $\ry^-(\t-\s)$ related to the 
 open-string ones  by \rf{68}, i.e.   as      $\ry^\pm(\t)=\pa_\t \rx^\pm(\t)$.
 In other words, given a   small closed string data  $\ry^\pm_i (\s)$    and replacing $\s \lar \t$ we  get 
 the corresponding  wavy line   data  $\rx^\pm_i(\t) $      according to \rf{68}.
 In the particular case we are interested in when  the open-string world sheet  is periodic in $\tau$ 
 with period $\Sigma$ \rf{spe}  the 
 %(we omit the   trivial smallness factor $\eps$ in \rf{68}). 
    total energy change of the open  string  should be computed over one period, giving 
\beq
   \Delta \bar{E}^{\rm op} = \bar{E}^{\rm op} ( \Sigma )-\bar{E}^{\rm op} (0) = \int_{0}^{{ \Sigma}}\!\!\! d\t\  (a^-{}^2-a^+{}^2) 
    =  \int_{0}^{{\Sigma}}   \!\!\! d\t\ \left[ (\pa_\t \ry_i^-)^2 -(\pa_\t \ry_i^+)^2 \right] \ ,\la{e13}
\eeq
i.e.  it  vanishes   once one  uses  the closed-string data  for $\ry^\pm$ 
satisfying    the level matching condition \rf{bb6}.

On the other hand,  the energy of the closed string  computed   directly  is given by
%22
\foot{Recall that $\t$ and $\s$ 
are interchanged in the two pictures, so  we replaced the  usual $\s$ integral by the $\t$ integral. } 
\beq
 \bar{E}^{\rm cl}
 %_{\mbox{closed}}
  =     \int_{0}^{{ \Sigma}}    \!\! d\t\ \left[ (\pa_\t \ry_i^-)^2 + (\pa_\t \ry_i^+)^2 \right] \ , \la{e133}
\eeq
which  in the dual (open-string)  language    can be written as
\beq
   \bar{E}^{\rm cl}
   %_{\mbox{closed}}
    =  \int_{0}^{{ \Sigma}}  \!\!\! d\t\  (a^-{}^2+a^+{}^2) \ , \la{e14}
\eeq
with the two terms contributing equally due to the closed-string constraint  \rf{bb6} (or the vanishing of  \rf{e13}). 

% Therefore, given a Wilson loop with boundary condition given by the shape and third derivative, near the boundary one splits the energy flux into incoming and outgoing. The sum of the two gives the energy of the dual short string. The difference, on the other hand, should integrate to zero which provides a
% global constraint on the possible boundary conditions. 

\def \bS  {\bar \S}

%%%%%%%%%%%%%%%%%%%%%%
%\subsection{Spin}

  Let us  also  comment    on  a similar  comparison    between   the    values of the  spin  of the  small  closed   string 
  and  the long open string. 
  For the closed  string in flat space the transverse spin components  are (in the  light-cone gauge, and suppressing the string tension factor as in 
  \rf{219}) 
\beq
 \bS^{\rm cl}_{ij} = \int^\Sigma_0  d\s (y_i \pa_\t y_j -  y_j \pa_\t y_i )\ . \la{353}
\eeq
 For the solutions $y_i(\t,\s) = \ry^+(\t+\s) + \ry^-(\t-\s)$ this   reduces, after integration by parts, to
\beq 
 \bS^{\rm cl}_{ij} =  \int d\s \left[ (\ry^+_i \pa_\t \ry^+_j -  \ry^+_j \pa_\t \ry^+_i ) -(\ry^-_i \pa_\t \ry^-_j -  \ry^-_j \pa_\t \ry^-_i )  \right]
        \equiv  \S^+_{ij} - \S^{-}_{ij} \ . \la{354}
\eeq
For the  near-straight open string surface in \rf{76}    we  may    define the  angular momentum components as 
\beq
 {\bS}_{ij} =  {\bS}^{\rm op}_{ij} = \int \frac{d\s}{\s^2} (x_i \pa_\t x_j -  x_j \pa_\t x_i ) \ . \la{355} 
\eeq
 However,  this  quantity  is not conserved, since the angular momentum can flow into/from the boundary: 
\be
\pa_\t {\bS}_{ij} = -\frac{1}{\s^2}\left( x_i\pa_\s x_j -x_j \pa_\s x_i \right) \Big|_{\s=\epsilon} =
  \frac{1}{\epsilon} \pa_\t (\rx_iv_j-\rx_jv_i) - \ha (\rx_i\pa_\s^3x_j-\rx_j\pa_\s^3x_i) (\t, 0)\ .   \la{633}
\ee
 This gives the influx of the angular momentum into the string in terms of the boundary data. 
%It is also interesting to use as boundary date the left and right moving functions $\rx^{\pm}_i$ in terms of which we obtain, up to total derivatives,
Expressed in terms of  the $\del_\t$ derivatives of the 
 boundary functions   $\rx^{\pm}_i$ this  reads (up to a total derivative) 
\beq
 \pa_\t {\bS}_{ij} = (v_i^+ a_j^+ - v_j^+ a_i^+)-(v_i^- a_j^- - v_j^- a_i^-) \ . \la{abv}
\eeq
Then the  total spin change   over one period  is  (cf. \rf{e13}) 
\beq
 \Delta {\bS}_{ij} = \int ^\Sigma_0 d\t \left[(v_i^+ a_j^+ - v_j^+ a_i^+)-(v_i^- a_j^- - v_j^- a_i^-)\right] = \S^+_{ij} - \S^-_{ij} \ , \la{356} 
\eeq
i.e.  it  is found to be equal to \rf{354}  after 
 we use again \rf{68}, i.e.   that  in going from the  closed  string  to  the open string picture 
the position  becomes  the velocity and the velocity becomes the acceleration.
%maps into position and the acceleration into the  velocity. % when going from the Wilson loop to the short string. 
The conclusion  is thus  that the total change in spin from the initial to the final state of the open 
 string equals the spin of the T-dual   short closed string.
 \iffalse
 \foot{
 {{\bf not sure   we need  remarks below, they are not very clear}
 Note that for the energy this was  not the case   because   of the  sign change for the left movers. 
 This is due to the fact that we have 
 T-dualized the current associated to the energy but not to the spin.
 %\foot footnote: 
 {As  was mentioned earlier, there are  other components of the spin, namely, 
 $\S_{4i}$,  which are  interpreted in terms of momentum and special conformal generator. Those are harder to interpret in the field theory and
 therefore we   will  not compute them here.}  }}
 \fi

 %%%%%%%%%%%%%%%%%%%%%%%%%%%%%%%%%%%%%
\subsection{Including  string motion  in $S ^5$}

%{\bf not sure this section belongs here} 

\def \rz {{\rm z}} 

%%%%%%%%%%%%%%%%%%%%%%%%%%%%%%%%%%%%%%%%%%%%%%%%
Let us now    consider the  case when the small string may move not only in $AdS_5$ but also in $S^5$.
We shall parametrize $S^5$  by the embedding coordinates as 
%comment also on  the spin in $S^5$. 
% In principle the short string can also have spin on the $S^5$. In the previous sections this was not considered since adding rotation on the $S^5$ 
% is a simple extension that we do in this section.  The sphere is given in embedding coordinates by
\beq
   Z_A Z^A +Z_6^2=1 , \ \ \ \ \ \ \ \ \ \ \ \ \ \ \ \ \  A=1, ...,  5  \ . 
\eeq
Like in  the $AdS_5$  case   in \rf{nei}, the small string limit  may be defined as  the following  expansion
% nea
% We explicitly separated $Z_6$ since we are going to expand around the point $X_6=1$, $X_A=0$. Consider then 
\beq
 Z_6 \sim 1 + \cO(\epsilon^2), \ \ \ \ \ \ \ \ \ \  Z_A\sim \epsilon \ . 
\eeq 
Then   the corresponding  $SO(6)$   currents are of order  (cf. \rf{214}) 
\beq
 j_{a,6A} \simeq \pa_a Z_A \sim \epsilon,  \ \ \ \  \ \ \ \ j_{a,AB}=Z_A\pa_a Z_B - Z_B \pa_a Z_A\sim \epsilon^2 \ . \la{361}
\eeq
 Since neither the  the T-duality \rf{tt}   nor the boost \rf{bu}  affect these coordinates, the same scaling 
  will apply also to the %would be true in the Wilson loop 
   open string   interpretation.\foot{Let us note that  one  may formally consider   similar T-duality relations between solutions 
 in $S^5$  directions. Indeed,  by complexifying the $S^5$  coordinates one  may  put it  into  the Euclidean 
 \ads (or de Sitter) form,  then 
 introduce  effective 
 Poincare  coordinates   with 4 linear commuting isometries  and  finally  apply the T-duality  \ci{rtw,bm}.
 However, this  prescription will in general map  real solutions   into complex ones  so its relevance
 in the present   context   is 
 not immediately clear.} 

 The short string in $AdS_5 \times S^5$ is described by the 8=3+5   transverse coordinates 
that, to the leading order,  solve  the  flat-space string equations,  i.e. in addition to 3  functions 
 $y_i$ in \rf{34},\rf{57} 
we get 5 extra  functions $z_A$  
\beq
   Z_A = \ep\, z_A (\s,\t) \ , \ \ \ \ \ \ \ \ \ \ \ \ \ \ \  z_A(\s,\t)=\rz_{A}(\s,\t) = \rz_A^+(\s+\t) + \rz_A^-(\s-\t)  \ . \la{362} 
\eeq 
In  what follows we shall again absorb $\ep$ into $z_A$.
%the short string, in light cone gauge  we just add five extra directions $y_{A}$. 
 
 The conformal constraints now  involve all  10 coordinates, so that in the light-cone gauge \rf{lc} 
 we get  relations generalizing \rf{bb5}  that   determine  $X_+(\s,\t)$  so that 
  the  level matching condition  \rf{bb6}  now is   % . Such level matching is now modified by the inclusion of the $S^5$ modes to 
\beq\la{lev}
   \int d\s\,  \left[ (\pa_\t \ry^+_i )^2 +(\pa_\t \rz^+_A)^2- (\pa_\t \ry^-_i )^2-(\pa_\t \rz^-_A)^2 \right] =0 \ . 
\eeq
 The small string  energy \rf{218}, \rf{e133} gets an additional contribution
\beq
 \bar E^{\rm cl} _{S^5} =  %\int_{-\infty}^{+\infty}\!\!\! d\t\ 
 \int d\s\, \left[ (\pa_\t \rz_A^-)^2 + (\pa_\t \rz_A^+)^2 \right] \ . \la{e133b}
\eeq
The $SO(6)$  charges $J_{6A}$ in the flat space limit become  momenta in these directions whereas
$J_{AB}$  represents the $SO(5)$  spins.  Let us  consider 
the   case  when  the total momenta $J_{6A}$ vanish. 
% and concentrate on the spins $J_{AB}$.
The expression 
for   the  $SO(5)$ spin  components    of the small   closed string  may  be  written as
\beq
 J^{\rm cl}_{AB} =  \int d\s\,  \left[ (\rz^+_A \pa_\t \rz^+_B -  \rz^+_B \pa_\t \rz^+_A ) -(\rz^-_A \pa_\t \rz^-_B -  \rz^-_B \pa_\t \rz^-_A )  \right]
         = J^+_{AB} - J^{-}_{AB}\ . \la{365}
\eeq
% These are the obvious flat space results for the short string. 
Let us  now turn to  the  open-string (Wilson loop)  picture. 
As  was already mentioned, the boost and the T-duality apply  to the 
$AdS_5$   coordinates only so the $S^5$  coordinates 
for the T-dual open string surface  are the same as for the small  string one.
On the gauge theory side the Wilson  loop is completely specified   by 
giving  the scalar   field  profile     parametrized  by the  5 small-fluctuation functions 
$ z^A$  as  $\Phi = \Phi_6 + \epsilon z^A(\t,\s=0)\Phi_A$.
% Consider now the Wilson loop. In fact, almost nothing has to be changed since the metric
% is still flat for the extra directions. The Wilson loop has small fluctuation in the boundary in the direction in which the scalar field $\Phi$ points. 
%It should be $\Phi = \Psi_6 + \epsilon y^A(\t,\s=0)\Phi_A$. 
This completely specifies the Wilson loop. However, in the   semiclassical    string theory     description we need  to 
fix   also  the derivative $\pa_\s z_A(\t,\s=0)$. 
Equivalently,  we  need  to  specify the functions  $\rz_A^+(\t)$ and $\rz^-_A(\t)$. 

 The energy and the angular momentum of the open  string  are  again 
  not conserved since there may be a  flow through the boundary. 
  The expression  for total  change of the open string energy 
   is a direct generalization of the one in \rf{e13} % in complete analogy with the previous calculation,
\beq
   \Delta \bar{E}= \bar{E}(+\infty)-\bar{E}(-\infty) = 
    \int_{-\infty}^{+\infty}\!\!\! d\t\ \left[ (\pa_\t \ry_i^-)^2+(\pa_\t \rz_A^-)^2 -(\pa_\t \ry_i^+)^2-(\pa_\t \rz_A^+)^2 \right]  \ , \la{e13b}
\eeq
and  as in \rf{e13} it   vanishes due to  the closed-string  level matching condition \rf{lev}. 
As  in \rf{e133}, the short  closed string energy is found  by adding the left and the right mode contributions. 
For the open string angular momentum we find the expression  similar to \rf{356}  
 \beq
  \Delta {J}_{AB} = J^+_{AB} - J^-_{AB}\ ,  \la{367} 
 \eeq
which is   again equal to  the closed string  one   \rf{365}.

%%%%%%%%%%%%%%%%%%%%%%%%%%%%%%%%%%%%%%%%%%%%%%%%%%%%%%%
\renewcommand{\theequation}{4.\arabic{equation}}
\setcounter{equation}{0}

 \section{Folded  spinning string case}
 
 The  folded spinning string \ci{gkp}  plays important role in gauge-string duality, with both long and short string limits
 providing important insights.  
 %a1
 Let us now  apply  the  above  T-duality  considerations in  this case without the assumption that the string is small.
 The T-duality maps this string to a Wilson loop with two boundaries, namely a heavy quark / anti-quark pair. At the end we may take
 the small string limit reproducing the result of previous sections plus corrections. In the open string language  this limit corresponds to zooming, e.g.,  
 near the quark and ignoring the anti-quark. 
 
 Our original motivation   was  to  understand a possible relation between  
 the  small-spin limit slope function \rf{4} and the near-straight line  Wilson loop   coefficient \rf{3}, 
 but we will not  try to address this question below. 
 
 We  shall  first consider the general spin case and then take  the small spin (short string) limit. 
 %and  compare to the  above general results. 
 % In this section we consider the T-dual of the GKP folded string \ci{gkp}  doing the flat space approximation. 
%  Since the boost can be done in 
%  the Poincare coordinates as a rescaling, the limit can be taken at the end and recover the previous results. 
   The     rigid spinning   string  ansatz   in \ads{5}  can be written   as 
   \be
    && X_{0}+i X_{-1} = \cosh\rho(\sigma)\, e^{i t(\tau)}, \ \ \ X_1+i X_2 = \sinh\rho(\sigma)\, e^{i \theta(\tau) },\ \ \ X_3=X_4=0\ , 
   \no \\   && 
    \rho=\rho(\sigma),\ \ \ \ \ \ \  \ \ \ t=\kappa\tau, \ \ \  \ \ \ \  \   \theta=\omega\tau\ . \la{an}
   \ee
   It represents%, in global coordinates, 
   a stretched rotating string with the center of mass fixed at the point $\rho=0$. The function $\rho(\sigma)$
   is given, in 
   the conformal gauge,   by any of the two equivalent relations 
   \beqa
    \cosh\rho(\sigma) &=& \dn(q\sigma,ik)\ , \ \ \ \ \ q\equiv \sqrt{\omega^2-\kappa^2} ,\ \  \ \ k\equiv \frac{\kappa}{q}\la{h}\ ,   \\ 
    \sinh\rho(\sigma) &=& k\ \sn(q\sigma,ik) \la{hh} \ . 
    \eeqa
    Here $\dn$ and $\sn$ are the standard Jacobi elliptic functions  (for our choice of their  definitions see Appendix A).\foot{Our notation here are different from the ones used, e.g., in 
    \ci{dun}  where  $k$ was $\kappa \ov \omega$, etc. The relation can be seen using 
    $$
    \cn(\omega\sigma+\mak,\frac{\kappa}{\omega}) =
        -\sqrt{1-\frac{\kappa^2}{\omega^2}} \frac{\sn(\omega\sigma,\frac{\kappa}{\omega})}{\dn(\omega\sigma,\frac{\kappa}{\omega})} = 
       - \sn(q\sigma,ik)\ , \  \ \  \ \ \ \ \ \ \ \  \mak= \mak({\kappa \ov \omega})\ , 
       $$
    which follow from %(we use relations and notation of
    \ 
   $
    \cn(u+\mak) = k' \frac{\sn u}{\dn u}, \ \ \sn(k'u,i\frac{k}{k'})=k'\frac{\sn(u,k)}{\dn(u,k)}, \  \  k' = \sqrt{ 1- k^2 }
   $ (see   \ci{grad}).
    }
   %These formulas are copied basically form tables in Grad, Ryzhik and use their notation (independent of our k, etc, \eg\ $k'=\sqrt{1-k^2}$)
   Below we will often  omit the modulus $ik$ from their argument in the assumption that it is the same for all Jacobi functions appearing in this paper. 
    %Other ways of writing the same functions are possible using equivalence relations between Jacobi functions. 
    
    The string is rotating with a physical angular velocity  $\Omega={ d \theta\ov d t}=\frac{\omega}{\kappa}$ and therefore only one of the parameters $\omega$, $\kappa$ is physical. Alternatively,  we can fix the
     periodicity of the coordinate $\sigma$ to be $\sigma\equiv\sigma+2\pi$ which gives
    \beq
    \frac{2}{q} \mak (ik)= 2\pi \ , 
    \eeq
   providing  a relation between $\kappa$ and $\omega$. 
   Note also that the  radial coordinate  $\rho$ varies from $0$ to its maximal  value $\rho_*$ given by 
   \begin{equation}
   \coth^2 \rho_* = \frac{\omega^2}{\kappa^2}\equiv 1+ \frac{1}{k^2} \ , 
   \end{equation}
   i.e. $k$ determines the length of the string.
   
   Using the definition of the Poincare coordinates in \rf{p1}--\rf{pok} we obtain
   the following   embedding  of the folded string solution into the Poincare patch of \ads{4}
\beqa
 && Z = \frac{1}{X_{-1}-X_4} = \frac{1}{\dn(q\sigma)\sin\kappa\t}\ ,\\
 &&   \X_0=\cot \kappa\t\ , \ \ \ \ \ \ \ 
  \ \ \ \X_1+i\X_2 = k\frac{\sn(q\sigma)}{\dn(q\sigma)} \frac{e^{i\omega\t}}{\sin\kappa\t}\ . \la{43}
\eeqa
%a1
% Here we suppressed the second argument of the Jacobi elliptic functions since it is always equal to the constant $ik$.
 Applying  the T-duality   transformation \rf{tt}  we get 
\beqa
  \tilde{Z} &=& \frac{1}{Z} =  \dn(q\sigma) \sin\kappa\t \ , \\
  \pa_\s\tilde{\X}_0 &=& -\frac{1}{Z^2} \pa_\t \X_0 = \kappa\, \dn^2(q\s)\ ,  \\
  \pa_\s (\tilde{\X}_1+i\tilde{\X}_2) &=& -\frac{1}{Z^2} \pa_\t (\X_1+i\X_2) 
   = k\, \sn(q\s)\, \dn(q\s)\, e^{i\omega\t} (\kappa\cos\kappa\t-i\omega\sin\kappa\t)  \la{45} 
  \eeqa
 These  equations  can be integrated (cf. \rf{aaa})   giving  the  explicit form of the  T-dual  solution:
 \beqa
  \tilde{Z} &=&  \dn(q\sigma)\sin\kappa\t \ ,  \\
  \tilde{\X}_0 &=& k\ \mathbb{E}(\am(q\s),ik)\ ,  \\
  \tilde{\X}_1 + i \tilde{\X}_2 &=& -\frac{k}{q}\cn(q\s)\, e^{i\omega\t} (\kappa\cos\kappa\t-i\omega\sin\kappa\t) \ . \la{46}
  \eeqa
  Interchanging $\sigma$ and $\tau$ coordinates 
   and dropping the tildes we find the final form of the open-string (Wilson loop)  solution 
\beqa
  Z &=&  \dn(q\t)\sin\kappa\s  \\
\X_0 &=& k\ \mathbb{E}(\am(q\t),ik) \\
\X_1 + i \X_2 &=& -\frac{k}{q}\cn(q\t)\, e^{i\omega\s} (\kappa\cos\kappa\s-i\omega\sin\kappa\s) \ . \la{49}
\eeqa
  The string ends at the boundary $Z=0$ on the two curves at
  $\s=0$ and $\s=\pi/\kappa$. They  are given by
\beq
 \X_0 = k\ \mathbb{E}(\am(q\t),ik)\ , \ \ \ \ \ \ \ \ \ \ \ \ \   \X_1 + i \X_2 = -k^2\cn(q\t)   \ , \la{47}
\eeq
and
\beq
 \X_0 = k\ \mathbb{E}(\am(q\t),ik)\ , \ \ \ \ \ \ \ \ \ \ \ \  \X_1 + i \X_2 = k^2\cn(q\t)\,   e^{i\pi\frac{\omega}{\kappa}}
 \ ,  \la{48}
\eeq
and thus  are   related by a simple spatial rotation.

 The energy of the closed string can be computed as the change in $\X_0$ over one period, namely
\beq
  \Delta \X_0 = \kappa \int d\t\,  \dn^2(q\t)\ .
\eeq
 The spin   can be  represented 
 as the boundary flux of the corresponding current. This can be seen generically by starting %before T-duality
 with the spin (${\rm S}={\rm S}_{12}$)  of the closed string given by (here $\X = \X_1+i\X_2$)
 \beq
  \bar {\rm S} = \int d\s\,  \frac{1}{Z^2} \Im (\bar{\X} \partial_\tau \X) = -\int d\s\,  \Im(\bar{\X} \pa_\s \tilde{\X} )\ ,
\eeq
where in the second equality we used the  T-duality relation \rf{tt}.
  Integrating by parts and using again the T-duality relation we obtain
\beq
 \bar  {\rm S}  = -\Im(\bar{\X}\tilde{\X})\Big|_{\s_i}^{\s_f} + \int d\s\,  \frac{1}{\tilde{Z}^2} \Im(\bar{\tilde{\X}}\pa_\t \tilde{\X})\ .
\eeq
  Interchanging $\sigma$ and $\tau$ and omitting  the tildes we find
\beq
 \bar  {\rm S}  = \int d\t\,  \frac{1}{Z^2} \Im(\bar{\X}\pa_\s \X)\ ,  
\eeq
where we also dropped the first term in view of the periodicity of the solution. 
It can be seen that  the spin of the open string vanishes, which means that the same
flux of the spin enters through one boundary and leaves through the other. 
This flux represents the spin of the closed string.

 The short string limit can be taken as $\kappa\rightarrow 0$ keeping $\omega$ fixed
 (so that $k\to 0$).  In this limit the solution  becomes essentially the  flat-space  folded  spinning string one 
\be 
t \to  k \tau  \ , \ \ \  \ \ \ \  \ \  X_1 + i X_2 \to   k \sin \sigma\  e^{i \tau} \ . 
\ee 
Explicitly,  we get in this limit
  \beqa
&&\sinh \rho = k \sin \sigma - \frac{k^3}{4}\sin \sigma\ \cos^2 \sigma +...  \ , \la{511}\\
&&\kappa=k \big(1 -\frac{k^2}{4}+...\big) \ ,\ \ \  \ \ \ \  \omega=1 + \frac{k^2}{4} + ...\ , \ \ \ \ 
q= 1 -\frac{k^2}{4}+...  \ , \la{512}
% \   \quad \rho'=\ \cos \sigma - \frac{\epsilon^3}{4}\cos^3 \sigma +...
% \ , \\
%&&\kappa \sinh \rho=\epsilon^2 \sin \sigma -\frac{\epsilon^4}{8}(3+ \cos 2 \sigma)\sin \sigma+...\ ,\\
%&& w \cosh \rho=  1+ \frac{\epsilon^2}{4}(1+ 2 \sin^2 \sigma)-\frac{\epsilon^4}{64}(8-\cos 4 \sigma)+... \ .
\eeqa
while   the spin   and the energy %(divided by $\sqrt \lambda$)
 have  the following   expansions  ($E= {\sql \ov 2 \pi} \bar {E}, \  \S= {\sql\ov 2\pi}  \bar{\S}$)
\bea
 \bar{\S}=  \pi k^2 (1 +  \frac{1}{8}  k^2 + ...) \ , 
%k=\sqrt{2 \mathcal{S}}\big( 1 -\frac{1}{8}\mathcal{S}+... \big) \ ,
\ \ \ \ \ \ \ \
\ \ \ \bar{E}=\sqrt{4\pi \bar{\S}}\Big( 1+\frac{3}{16\pi }\bar{\S}+...\Big)\ . 
\label{mfj} 
\eea
 The lowest order expression  for the corresponding T-dual solution \rf{49}  is then 
\beq
  Z = k \s\ ,\ \ \ \ \ \ \ 
   \ \ \X_0 = k  \t\ ,\ \ \ \ \ \
    \ \ \X_1+i\X_2 =  k^2 \cos \t \, e^{i \s} (1-i \s)
     %-\frac{\k^2}{\omega}\cos\omega\t \, e^{i\omega\s} (1-i\omega\s)
    \ . \la{539}
\eeq
 After the rescaling by $k$      which corresponding to the boost in \rf{bu}      we find \foot{Let us note  once  again
 that  in the limit of $\S\rightarrow 0$  the string is short  and thus should   be  close to a massless particle
moving with the speed of light. This, however, is not reflected in the present   solution which
 degenerates to 
$\r=0$ in this limit. 
% At the exact value $S=0$ the string would be massless and therefore it is unphysical that it is sitting at $\rho=0$, it should be 
One way to avoid this problem is to add   momentum $J$ in   $S^5$    so that in the limit $\S\to 0$ the string  solution reduces to the 
 BMN geodesic (representing  a massless  BPS state in the full  $AdS_5 \times S^5$  theory). 
An alternative  way  to ensure a non-trivial BPS limit  that we  consider here 
  is to generalise the  folded string  solution by  boosting it  within $AdS_5$ 
%This problem can be avoided if the string is boosted in such way
so  that the $\S\rightarrow 0$ coincides with the limit of infinite boost and  therefore 
 the resulting point-like  string  follows  along a massless geodesic.}   
\beq
 Z = \s\ , \ \ \ \ \ \ \ \ \X_0 = \t\ ,\ \ \ \ \ \    \ \ \X_1+i\X_2  = Y - \s\pa_\s Y\ ,    \la{50}
\eeq
 with $Y$ being   a  harmonic function 
 \beq
  Y \equiv - k  \cos \tau\, e^{i \s} \ . \la{514}
  % -\frac{\kappa}{\omega} \cos\omega\tau\, e^{i\omega\s} \ . \la{514}
 \eeq
  Notice that $\pa_\t Y = \sin\tau\, e^{i\s}$ which is indeed 
  the small string limit of $X_1+iX_2$ in eq.\rf{an} as expected form our previous analysis in section 3  (see \rf{61}--\rf{68}).
 
  It is easy to extend the small $\kappa$ (small $k$)  expansion to higher orders.
  Including the first subleading terms we  find  (using \rf{512})
\beqa
Z &=& \s-\frac{k^2}{6} \s (\s^2-3\sin^2 \t)\ ,  \ \ \ \ \ \ \ \ \ \ \ \ \ \ \ \ \ \ 
\X_0 =\t +\ha k^2 (\t-\sin\t \cos\t ) \ , \la{515}  \\
 \X_1+i\X_2 &=& -k \cos \t\, e^{i\s} (1-i\s) -\frac{ik^3}{12} e^{i\s} \Big[3 i \sin\t (\t-\sin\t\cos\t)(1-i\s)+2\s^3\cos\t\Big]\ .  \no
\eeqa
 Similar  discussion can be repeated for 
 the small string limit  of other closed string solutions in the bulk of \ads{}. We shall  consider 
 the  spiky string  case  in  Appendix  B.

 \def \w  {\omega}

\renewcommand{\theequation}{5.\arabic{equation}}

\setcounter{equation}{0}

%%%%%%%%%%%%%%%%%
\section{Comments on interpretation of the T-duality relation}

 Let us  now  make  few comments on a  physical interpretation of the  T-duality   relation
 discussed   above.
 
%\subsection{Comments on  interpretation of T-duality relation }
   The  Poincare  patch of the   \ads{} space 
     is the near horizon limit of the extremal D3 black brane   geometry, and as such, it has a  horizon at $Z=\infty$.
   In global   \ads{} coordinates in \rf{ep}  such  a horizon is a light-like surface given by $X_-=0$. 
 Let us   consider  a small string    in the bulk   which after a boost   
  falls into the horizon of the extremal black brane  at the point $X_0=1$, $X_\pm=X_i=0$. 
   After the T-duality, the string falling into the horizon becomes effectively  an open string
   with the boundary end-point   that may be interpreted as a 
    heavy ``quark'' 
 moving along a prescribed trajectory in the dual field theory.
 
 The quark's  trajectory is periodic with period $\Sigma= {\bar{E}}$
  where $\bar{E}$ is the closed-string energy (see \rf{spe},\rf{63}).
 The velocity of the quark is directly given \rf{68}
 by the position of the closed string when falling through the horizon. 
 The quark absorbs and radiates the energy and the spin.
% In the present AdS/CFT case 
The  energy 
 is expressed  \rf{e9} in terms of  the third derivative of the string position and is given by the momentum density of the string. 
 Loosely speaking, this  momentum density  determines the state of the field that surrounds the quark
 in the dual field theory picture.\foot{In the field theory, on a first approach, if we want to reproduce the energy vs. spin relation, we need to compute the relation between the energy and the spin of the radiation emitted by the quark. However,
  this should be integrated over trajectories. The precise computation to be done in the field theory is not clear and is left as important problem  for the future.} 
 
  %%%%%%%%%%%%%%%%%%%%%%%%%%%%%%%%%%%%%%%%%%%%%%%%%%%%
 For   a pictorial representation of  the  \ads{}  space  it is convenient to introduce a new coordinate $0<\xi<\frac{\pi}{2}$ %through\beq  \cosh\rho = \frac{1}{\cos\xi} \ ,  \eeq
     so that the  $AdS_5$    metric becomes (cf. \rf{51}) 
\beq
     ds^2 = \frac{1}{\cos^2\xi}\left(-dt^2+d\xi^2 + \sin^2\xi\ d\Omega^2_{[3]}\right)\ , \ \ \ \ \ \ \ \ \ \ \ \ \ \  \    {\cos\xi}  =    { 1 \ov  \cosh\rho} \ .   \la{667}
\eeq
The Poincare coordinates are chosen in such a way  that the horizon is at $X_-=0$. 
The boost performed  in \rf{57},\rf{59} squeezes the string into a small region around $X_+=0$
(see  fig. 1 and fig. 2).  %\ref{fig1}. 
%A10 %21
 There is, of course, an ambiguity in choosing the Poincare patch. For another choice, the horizon will cut the world-sheet of the 
 closed string on a different curve and therefore the dual Wilson loop will be different. This is expected since choosing a different Poincare
 patch is tantamount to choosing different directions to do 
 T-duality and therefore the result  may be  different.\foot{The T-duality the way we defined it in Minkowski signature 
 relies on a   choice of  the   Poincare patch. 
  One may  start instead   in the Euclidean signature \ads{} where the Poincare  coordinates cover the full space 
  and use also the Euclidean signature on the world sheet. Then the  T-duality will in general map real solutions to complex ones 
  and  one  will need to    use an analytic continuation to define  the T-duality map  back in Minkowski signature
  (though  this procedure may not, again, be unique).  In global (e.g., embedding) coordinates 
  the ambiguity in  the definition of the T-duality map will be reflected in an   ambiguity in a choice of 4   commuting isometries 
  used to perform the T-duality.  In the   Minkowski signature that   will  correspond to  having different horizons. 
 In each case  the T-dual  of a small string   will be  a
different Wilson loop corresponding to  ``freezing" the string at different times.
The ``crossing of the horizon'' picture   is  relevant    because the shape of the string at the horizon is what maps to
the Wilson loop shape, namely, the surface $X_- = 0$  maps to boundary. In the above discussion we put a short string
at $\rho=0$  so the crossing the horizon is unavoidable.}

 %%%%%%%%%%%%%% %%%%%%%%%%%%%%%%%%

\begin{figure}
\begin{center}
\includegraphics[width=14cm]{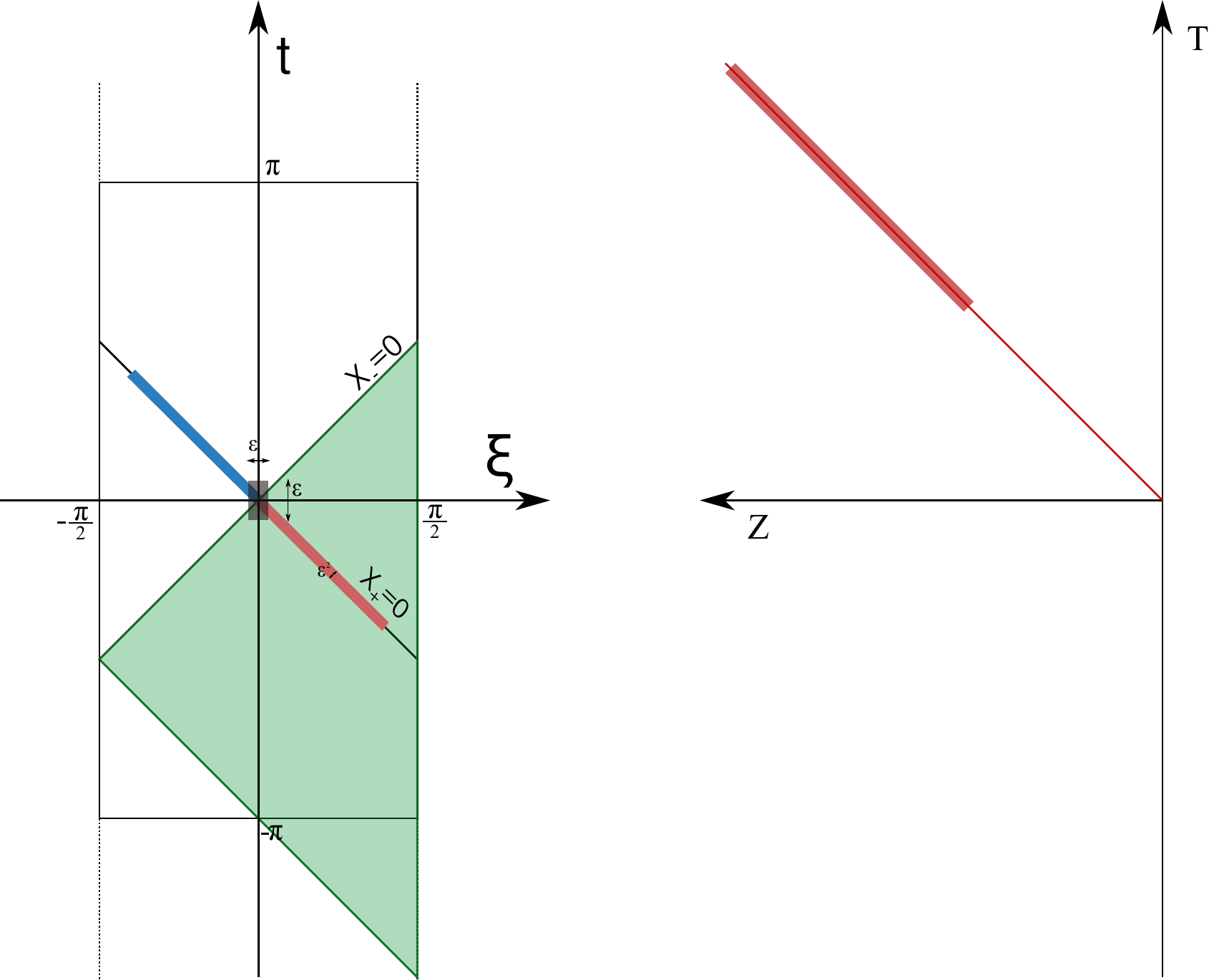}
\caption{\ads{} in global coordinates with the Poincare patch defined by the horizon $X_-=0$ shaded in green. We consider a short 
string moving in the gray region of linear size $\epsilon$ around $X_0=1,\  X_\pm=X_i=0$. Since the string in nearly massless, 
after a  boost  it moves in the red colored region of size $\epsilon^2$ around the light-like line $X_+=0$, 
 arriving at the horizon  at the speed of light. 
The state in which it falls through the horizon is given by its shape and momentum density and maps directly to the properties of the dual Wilson loop.
 The Poincare patch (before T-duality) is shown on the right. 
 The string is then almost point-like and moving in a neighborhood of the geodesic indicated in red towards the horizon at $Z=\infty$.}
\end{center}
\label{fig1}
\end{figure}

%%%%%%%%%%%%%%%%%%%%%%
\begin{figure}
\begin{center}
\includegraphics[width=14cm]{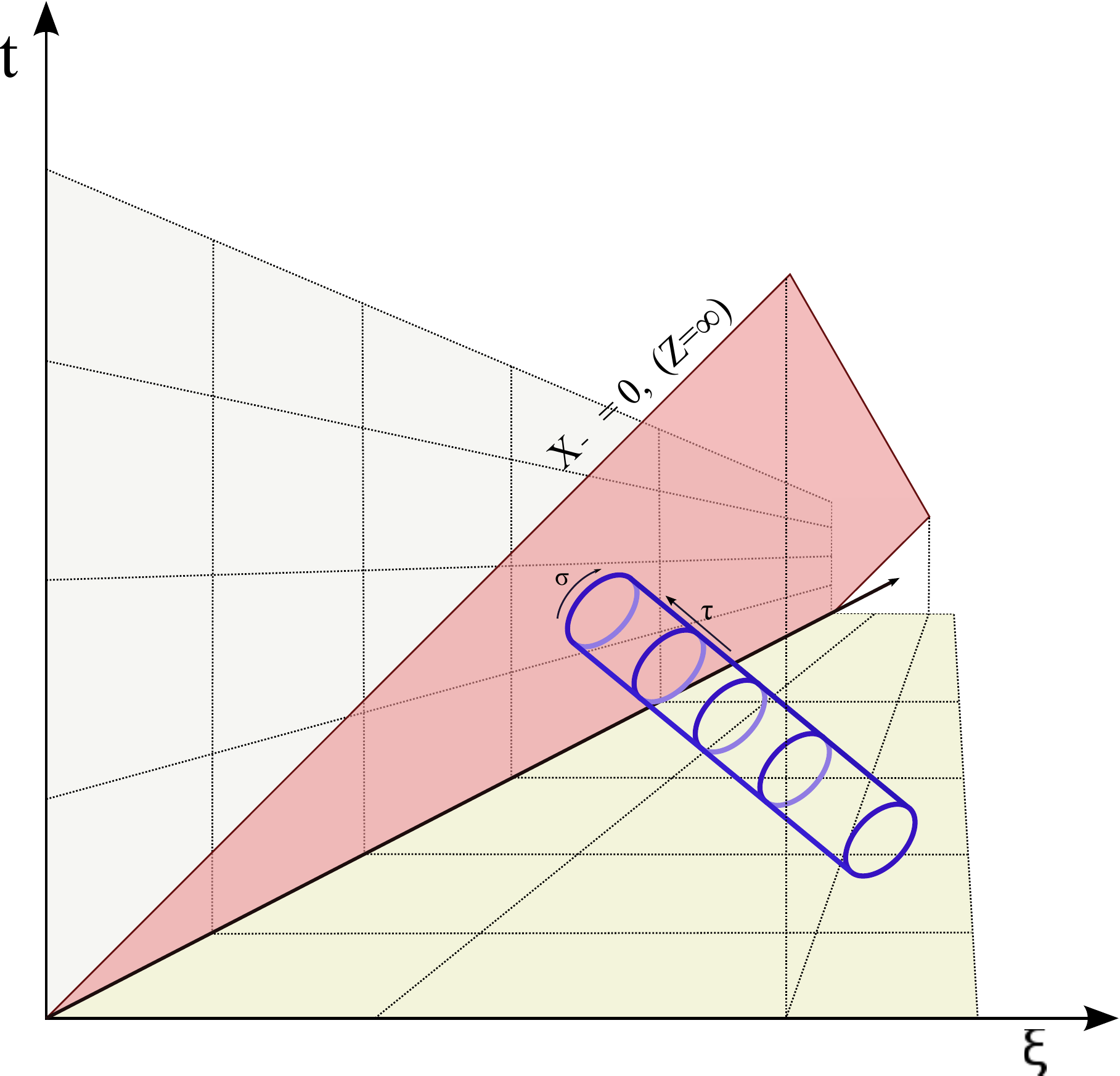}
\caption{%{\bf Figure probably not good for paper but I think it helps fix ideas for the moment.}\\ 
       %  After the boost  %we are in a situation where a 
         A  small boosted (nearly-massless) string % (near massless)
          falling into the extremal horizon. 
          Let  
           $j_{\s,\t }$  be  the components of  the current associated with the energy conservation
           (here $(\t,\s)$   are the  world-sheet   coordinates).
       %   Using $\t$ as in figure one can 
          We may compute  the energy of the string as an  integral in $\sigma $ right at the horizon $ \int d\s\, j_\t$. Interchanging $\s$ and $\t$, 
           we  get  a string with a  boundary and a flux along the end. The same integral  %as before (giving the same result) 
           $\int d\t\, j_\s$
           can be  interpreted as a flux of energy leaving the string through the boundary.
           The  two expressions  are the same since  they are given by the same integral. 
       It is the   second interpretation that applies       after the T-duality. % second interpretation applies. 
       }
\end{center}
\label{fig2}
\end{figure}

 At the boundary there is an incoming and outgoing  flux of energy.
The difference between the fluxes gives the (non-conserved) energy on the open string (Wilson loop) side (see \rf{e11},\rf{e13}).
On the other hand,  the total integrated flux (given by the sum of  incoming plus outgoing fluxes) gives the energy of the dual closed string \rf{e133}.
For periodic closed string the total integrated incoming and outgoing fluxes are the same (due to level matching) so they contribute
equally to the energy.
%T-duality interpretation here is natural  since level matching is actually wrapping. 
%Therefore they get interchanged with energy.
The   parameter   entering the level matching condition  (``wrapping'')
 is a difference in the  momentum flux, and the energy is the sum.
  They get interchanged  after  the  T-dulaity.

% OK. I think we can say this in a funny (interesting) way.  

In general, the T-duality  is a  kind of   ``position $\leftrightarrow$  momentum'' duality (cf. \ci{am}). 
Here  we considered   another  example making  it  explicit that the local degrees of freedom (strings) near the horizon 
are those of the dual momentum theory.  A natural expectation  is % assumption 
% Now we add the information 
 that  a string % object 
 that falls through the horizon  should correspond  to  an 
insertion of an operator (Wilson loop) in the T-dual theory.  Here we have given   a  precise map
  for short strings: for a closed string that 
crosses the horizon we have to insert a Wilson loop  with the 
corresponding shape 
determined by     the shape of the closed string. % in the momentum theory.

 %Let me say it in a more explicit way. S
More explicitly, suppose we have a process where we send two strings from 
the boundary (represented by  dual operator insertions) which  collide in such a way that one outgoing string goes to 
the boundary (another operator insertion) and the other goes onto the horizon. 
Then  in the dual field theory  one has to insert a Wilson loop to represent the  latter  string. 
This  Wilson loop is precisely the one defined by a wavy line that  we  discussed. 
Unfortunately,  we do  not  know  at present how to 
write such a Wilson loop  explicitly in terms of  the original open string  variables,  % but that's another problem. 
though  we have a small closed string description for it.\foot{Note that   in \ci{sevv,druu}
 there is a similar relation  between closed/open string descriptions based on integrability (reflection matrix).   
  In Minkowski case  there should be more freedom.}

% There are few  extensions and open problems. 
% One is connection  to operators  and Wilson loops    in the context of relation  between their correlators  at null separation. 
 %Relation between  exact results for slope small strings and    wavy Wilson lines. 
 
%Here we discussed a  map  between position of short string and velocity of  trajectory defining the Wilson loop. 
%Therefore, a spiky string that has cusps or  jumps  in its  shape (in $\s$-``velocity"), corresponds to Wilson loop with jumps
%in acceleration (but no cusps).
%Vice versa, a Wilson loop with cusps has jumps in velocity and therefore  should  correspond  to string with jumps in ``position'', 
%reminding jumps in momentum in  the open string picture in \ci{am}. 
%which does not make sense.
% Although the correct interpretation of jumps in position seems to be open strings   as
%per Alday-Maldacena idea.
% Probably we can work out WL for spiky string since spiky string in flat space is quite trivial. 
%It is  interesting to work out 
%Some comments on the spiky string case %example in more detail. 
%are made in the Appendix B.

%%%%%%%%%%%%%%%%%%%%%%%%%
  \begin{figure}
\begin{center}
\includegraphics[width=14cm]{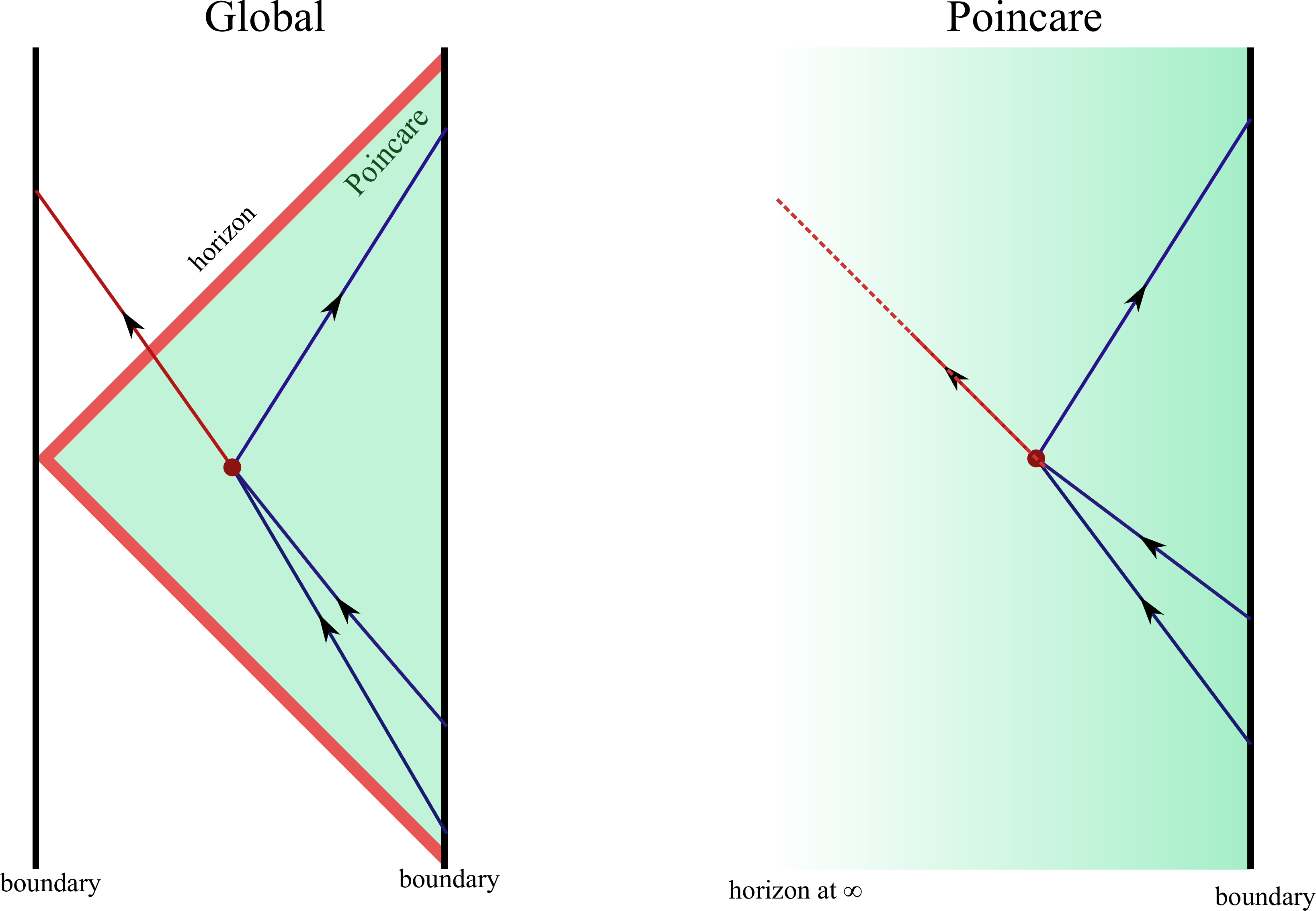}
\caption{An amplitude  with three insertions at  the boundary of 
the Poincare patch and one on the other boundary. 
The  Poincare patch of the  \ads{} space  in global coordinates is    shaded in green. 
In global coordinates this  amplitude  is computed by a
 world-sheet  theory  with four local operators inserted.
In the Poincare patch, the world sheet has a boundary. 
The boundary conditions, in the T-dual interpretation, are exactly those of a world
sheet ending at the boundary of \ads{} space on a line (Wilson loop)
 with shape and third derivative given by the state of the string falling in.
  Therefore, from the boundary theory point of view, 
 the amplitude is   given by a ``mixed''  correlator --  with three local operator insertions 
  and a non-local Wilson loop of the T-dual ``momentum'' theory. Although it is not easy to compute it, 
  it is interesting that at least in principle the dual  \N{4}  gauge theory is self-contained. This  is reminiscent of a black hole complementarity but here we are
not trying to describe what happens behind the horizon. }

\end{center}
\label{fig3}
\end{figure}
%%%%%%%%%%%%%%%%%%%

 % Let us   add   some further comments on T-duality and  scattering amplitudes. 
 In the Euclidean AdS/CFT correspondence, \N{4} Euclidean gauge invariant correlators can be computed as string theory correlators in Euclidean \ads{} for strings that emerge from the boundary by the insertion of appropriate vertex operators. If we consider the Minkowski signature case in global coordinates, 
 the same correspondence applies   (see fig. 3).
  However,  when selecting a Poincare patch, it may  happen that one (or more) of the local operators inserted in not
 on the boundary patch, or, equivalently,  one (or more) of the closed strings falls into the extremal Poincare horizon.
  %From the  string theory  point of view, 
 Then the world sheet  has a boundary: 
 in the example of  fig. 3  the world sheet has three operator insertions plus a boundary at the horizon. 
  In this paper we saw that the  conditions at such boundary can be simply expressed 
  by  saying that the world sheet ends at a wavy line  Wilson loop 
in the T-dual version of the  \ads{} space. 

Note  that the same \ads{} space can be interpreted as
corresponding to  the original \N{4} supersymmetric gauge  theory or to the T-dual (``momentum'') theory.
Therefore, from the  dual gauge theory  point of view we can compute the
 same amplitude by replacing the missing local insertion by a Wilson loop of the T-dual 
``momentum'' theory. 

The situation is somewhat reminiscent of the black-hole complementarity,
 although it is not obvious that  there is any information about what 
happens behind the horizon. It should be noted  that  the present situation corresponds to 
 an (extremal) eternal black hole and therefore the usual paradoxes due 
to black hole evaporation do not arise. Nevertheless,  it is interesting to note that the correlation
function can be computed entirely in terms of the operators in a single \N{4} gauge theory. 
The computation is not straight-forward since the dual  (``momentum'')  Wilson
loop is not easy to represent in terms of the original  variables. 
This is again similar to the complementarity where the missing local operator is scrambled 
in a non-trivial way when represented in the original variables.  
Note also  that we insert an operator related to the state of the
 string when crossing the horizon and not directly related to the operator on the other boundary. 

%\subsection{Comments on scattering amplitudes}

\def \Si {\Sigma}

  %%%%%%%%%%%%%%%%%%%%%%%%%%%%%%%%%%%%%%%%%%%%%%%%%%%%%

\renewcommand{\theequation}{6.\arabic{equation}}

\setcounter{equation}{0}

 \section{Discussion}

 The main result of this   paper is 
  the demonstration of   existence of a map between the state of a closed string falling into the Poincare horizon and a Wilson loop 
of the T-dual boundary theory. 
The  T-duality  we used  is like the one of \ci{am}  that relates  gluon  scattering %open string 
amplitudes  
to closed light-like Wilson loops with cusps. 
In our case the  profile  of  the corresponding Wilson loop is   completely determined
 by  the  shape of the closed string. In particular, it is  periodic with period  fixed  by   the energy 
of the closed string. 
 This allows us  to %precisely
define correlation functions of local operators where one  of the corresponding 
strings falls into the horizon. The string that  crosses the horizon
 is replaced by a Wilson loop of the T-dual  theory. 
 %Although such mixed correlator is not easy to compute,  it is, in principle,  well defined and gives a prescription 
 %of  how to describe the Poincare 
%horizon in the field theory. 

%%%%%%%%%%%%%%
 Our semi-classical computation has barely scratched the surface of what seems 
 to be an interesting topic  for  further research. 
 For example,  these  ideas  might have  implications for understanding the physics of black holes, e.g.,  black hole complementarity. 
 For example, it suggests that the dynamics of closed strings in a small  flat-space region in the bulk of \ads{} 
 %region which can be approximated by flat space 
 is dual to  a  correlation
 functions of certain Wilson loops in  gauge theory.
 % the case when 
 % all closed strings  emerge from 
% or fall into the horizon  may be  described
% by Wilson loops correlators of the T-dial theory which is 
%  same as the original one  (with  no
 % correlators  involving operators   theory and the T-dual are involved. 
 % (in particular, in connection with some recent work 
%It might be useful to understand better recent ideas such as those in
% \cite{BH1,BH2}).
 To make this precise the first step   would be to describe the T-duality map at the full quantum level.
  The algebra of oscillators of the closed string
 should map to the algebra of the corresponding deformation operators of the Wilson loop. 
 Also,  the partition functions of the T-dual world-sheet theories should map to each other, 
implying that the relation   holds  to all orders in $\frac{1}{\sqrt{\lambda}}$ expansion.
% Although apparently simple to do, it seems involved to put it beyond the
%scope of this paper and therefore is left for future work. 
%Initial steps towards understanding
%Related comments are 
                %Comments  on  t  relation between the partition functions  are 
%made in 
%  Appendix  C. 
% where  we  briefly discuss %this point at the bosonic level with the main idea of 
%understanding
 %how the T-duality works for the cylinder representing a closed string insertion

The T-duality relation discussed  above 
 implies that   one should be able to translate results  known about Wilson loops   into statements about  
closed strings crossing the horizon. 
%AT
 In particular, the exact result  for the wavy-line  open string energy \rf{vv} 
 should   have its counterpart on the small closed string side, potentially providing a link to the slope function in \rf{4}. 

%%%%%%%%%%%%%%%
 On  the  dual field theory side,  the  corresponding  problem is how 
  to  compute  a   relation between the spin and the energy radiated by a heavy quark
  moving along a wavy line trajectory. 
 At lowest order   one can phrase this question 
 in the language of \cite{cor}. %For example,
  If the period is $\Si$,  %the trajectory can have a wave number $n$, namely 
  then the  quark oscillates
with frequency $\omega = \frac{n}{\Si}$  
($n$ is a  wave number). Let  the  amplitude of oscillations  be $\eta$. 
According to \cite{cor} the quark radiates   quanta of energy $\omega$ with
probability 
\beq
 p_{_{\mbox{emission}}} =   f(\lambda)\, \omega^3 |\eta|^2\Si \ . \la{ra}
\eeq
where $f(\lambda)$ is a known function of $\lambda$  scaling as  $\sql$ at strong coupling.
 If those quanta in addition to  energy $\omega$ carry    spin $1$ (gluons) 
 then  the energy and the spin radiated  by the quark are
\beq
   E = f(\lambda) \omega^4|\eta|^2 \Si  \ , \ \ \ \ \ \ \ \ \ \ \ \ \ \ \ \S = f(\lambda)  \omega^3 |\eta|^2\Si \   . \la{rre} 
\eeq
It  follows then that 
\beq
  E = \omega \S = \frac{n}{\Si} \S\ . \la{ree}
\eeq
%Notice that the presence of the  $\sqrt{\lambda}$
% factors  in  \rf{ra},\rf{rre} follows  from the strong coupling limit of the exact expression  of $\cite{cor}$ and 
%is a highly non-trivial result. 
Let us now try to interpret this relation in T-dual  closed string terms. 
As we discussed  above, 
 the period $\Sigma$   should  be proportional  \rf{spe} 
 to the  closed-string energy  $\Si %=\bar E  = \frac{2\pi E}{\sqrt{\lambda}}$.
 \sim \frac{ E}{\sqrt{\lambda}}$.
 For the open string T-dual to closed string  there should be both left and right mover  contributions,   and 
% Although we discussed emission, 
 the level matching condition  implies  that the
  left and right movers emit and absorb the same total energy.
Then  it follows from   \rf{ree}   that  % and therefore
% ({\bf SEEMS here    E open and E closed got identified ?})
 \beq
  E \sim  \sqrt{ {\sql\, n \S}}\  .%\ \ \ \ \mbox{with} \ \ \ \alpha'=\frac{1}{\sqrt{\lambda}}. \la{eer} 
 \eeq
 This is  a Regge trajectory relation 
 % with slope $\lambda^{{1}/{4}}\sqrt{n}$ 
 for a closed string state  at level $N=\tilde{N}=\half n S$ and tension $\sim \sql$.  
% {\bf HOW do you know that ?  why not a state that satisfies level matching ?} 
% Such a  state does not satisfy the level matching 
% condition.
%  Including the left movers corresponds to including the 
%  absorption by the quark (cf. \rf{e11},\rf{e13},\rf{e14}). 
%Assuming that the spin flux into the open string is maximized, we simply have that the total energy and spin duplicate but the formula
%$E=\omega S$ still holds and the Regge trajectory does not change. 
 This naive computation should, of course,  be made
 more precise. It is important to understand,  for example, how the $1 \ov \sql$ corrections appear. 
 
 % It is clear that there are many directions to further explore and we leave that for future work.  

%%%%%%%%%%%%%%%%%%%%%%%%%%%%%%%%%%%%%%%%%%%%%%%%%%%%%
\section*{Acknowledgements}

We would like to thank  A. Mikhailov and  R. Roiban  for  very useful discussions. 
 The work of AAT was supported by the  STFC grant ST/J000353/1
and by the ERC Advanced grant No.290456. 
 The work of M.K. was supported in part by NSF through a CAREER Award PHY-0952630, and by DOE through grant DE-FG02-91ER40681.

%%%%%%%%%%%%%%%%%%%%%%%%%
%\appendix
%\addcontentsline{toc}{section}{Appendices}
%\addcontentsline{toc}{section}{Appendices}

\appendix
%\numberwithin{equation}{section}
%\renewcommand{\theequation}{A.\arabic{equation}}
%\setcounter{equation}{0}

%\refstepcounter{section}

\section{Elliptic Integrals  and Jacobi elliptic functions
}
%\refstepcounter{section}
\def\theequation{A.\arabic{equation}}
\setcounter{equation}{0}
%\section{Appendix: Symmetries}

The rotating string solution and its T-dual are written in terms of Elliptic integrals and Jacobi Elliptic functions. Since there are different
notations and definitions used for these functions, especially in the case of an imaginary modulus, in this appendix we briefly collect the 
definitions that we use, together with some simple properties of these functions.
 We use the following notation for elliptic integral $\mae$
\beq
\mae(\phi,ik) = \int_0^\phi d\theta\, \sqrt{1+k^2\sin^2\theta} \,  \ . 
\eeq
Notice that with this definition  $\mae$ is an always increasing function; in particular,  it is not periodic. This is the 
appropriate definition for an  imaginary modulus. 
Also we define 
\beq
\sn(z,ik) = \sin \phi(z,ik), \ \  \cn(z,ik) = \cos\phi(z,ik), \ \  \dn(z,ik)=\partial_z \phi(z,ik)\ , 
\eeq
where $\phi(z,ik)$ is determined by 
\beq
z = \int_0^{\phi(z,ik)} \frac{d\theta}{\sqrt{1+k^2\sin^2\theta}} \ . \la{aa}
\label{phidef}
\eeq
Again, for imaginary modulus,  $\phi(z,ik)$ is an arbitrary real number and is not a periodic function of $z$. On the other hand, the functions $\sn(z,ik)$, $\cn(z,ik)$ and $\dn(z,ik)$ are periodic.  
Taking $z$ derivatives on both sides of eq.(\ref{phidef})  gives 
\beq
 %1 = \frac{\partial_z \phi(z,ik)}{\sqrt{1+k^2\sin^2\phi(z,ik)}}  \ \ \Rightarrow \ \ 
 \partial_z \phi(z,ik)=\sqrt{1+k^2\sin^2\phi(z,ik)} \ .
\eeq
 It follows that 
 \beqa
 \partial_z \mae\big(\phi(z,ik),ik\big) &=& \partial_z \phi(z,ik) \sqrt{1+k^2\sin^2\phi(z,ik)}\no  \\
                                &=& 1+k^2\sin^2\phi(z,ik) = 1+ k^2 \sn^2(z)\ . 
 \eeqa
 This  implies the  following expression for the integral
 \beq
 \int\! d\sigma \ \sn^2(q\sigma,ik)  = -\frac{\sigma}{k^2} +\frac{1}{q k^2} \mae(\sn(q\sigma),ik) \ .   \la{aaa} 
 \eeq
  %used in the main text.
 
 %%%%%%%%%%%%%%%%%%%%%%%%%%%%%%%%%%%%%%

 \section{Spiky string case}
 %{\bf For reference only, may be we eliminate this appendix later. I'll include calculations for reference here.}
 
 %\refstepcounter{section}
\def\theequation{B.\arabic{equation}}
\setcounter{equation}{0}

 %\section{Spiky strings}
 %%%%%%%%%%%%%%%%%%%%%%%%%%%%%%%%%%
 Let us consider T-dual of spiky string in  \ads{} space  \cite{spiky}. 
 The corresponding 
 flat space solution in the conformal gauge  is\ 
\beqa
  y_0&=& \t  \ ,\ \ \ \ \ \ \ \    \ \ \ \ \ \ \  \ \ \ \ \ \ \    y_{3} =y_4= 0\ , \no  \\
  y_1 &=& \kappa\Big\{\cos \Big[\frac{\t+\s}{2\kappa}\Big] + (n-1) \cos\Big[\frac{\t-\s}{2\kappa(n-1)}\Big] \Big\}\ ,  \no \\
  y_2 &=& \kappa\Big\{\sin \Big[\frac{\t+\s}{2\kappa}\Big] + (n-1) \sin\Big[\frac{\t-\s}{2\kappa(n-1)}\Big] \Big\}   
  \ . 
\eeqa
%{\bf amplitude should be independent of $\kappa$-- we may  set $\kappa= 1/2$ and keep amplitude any}
Here  we  choose $y_0=\tau$   by  rescaling of $\tau$ and $\sigma$, so that  $\s\equiv\s+4\pi\kappa(n-1)$
where $n$ is the number of spikes  % and $\s\equiv\s+4\pi\kappa(n-1)$
($n=2$ is the usual folded spinning string).
%This represents the  flat-space limit of the spiky string solution in 
  %\ads{} space  \cite{spiky}. 
  %For $n=2$ the flat space solution is the usual folded rotating string.
  %and its   \ads{} counterpart is the  \cite{gkp}.
 
 This   may be viewed as solution  in  $R^{1,4}$  in the  
  light-cone gauge if we take the advantage of the fact that $y_4=0$ and define $y_\pm=y_0 \pm y_4 =\t$
(cf.  \rf{59}).\foot{A different choice, for example $y_\pm=y_0 \pm y_1$, leads to complications which 
%can be best  avoided for the moment but which 
need to be dealt with if $y_{3,4}\neq 0$.} 
 The left and right moving parts of  $y_i = \ry_i^+(\t+\s) + \ry_i^-(\t-\s)$  are easily identified as
\beqa
&& \ry_1^+(\t) =  \kappa \cos \Big[\frac{\t}{2\kappa}\Big],\ \ \ \ \ \ \ \   \ry^-_1(\t) =\kappa (n-1) \cos\Big[\frac{\t}{2\kappa(n-1)}\Big] 
\ , \\
 &&\ry_2^+(\t) =  \kappa \sin \Big[\frac{\t}{2\kappa}\Big],\ \ \ \  \ \ \ \  \ry^-_2(\t) 
 =\kappa (n-1) \sin\Big[\frac{\t}{2\kappa(n-1)}\Big] \ . 
\eeqa
%so that $y_i = \ry_i^+(\t+\s) + \ry_i^-(\t-\s)$. 
The corresponding open string (Wilson loop)
 data  follows from the general expression for the T-dual solution  in section 3 (see \rf{68}). 
After the  redefinition $\s\leftrightarrow \t$ we get 
\beqa
 \T&=&\t  \ ,  \\ 
 \Y_1 &=& 2\kappa \epsilon \Big\{  \sin \Big[\frac{\t}{2\kappa}\Big] + (n-1)^2  \sin\Big[\frac{\t}{2\kappa(n-1)}\Big]  \Big\} \ ,  \\
 \Y_2 &=& - 2\kappa \epsilon \Big\{  \cos \Big[\frac{\t}{2\kappa}\Big] - (n-1)^2  \cos\Big[\frac{\t}{2\kappa(n-1)}\Big]  \Big\} \ ,  \\
 \pa_\s^3 \Y_1 &=& \frac{\epsilon}{2\kappa} \Big\{  \cos \Big[\frac{\t}{2\kappa}\Big] - \frac{1}{n-1} \cos\Big[\frac{\t}{2\kappa(n-1)}\Big]  \Big\}  \ , \\
 \pa_\s^3 \Y_2 &=&  \frac{\epsilon}{2\kappa} \Big\{  \sin \Big[\frac{\t}{2\kappa}\Big] + \frac{1}{n-1}  \sin\Big[\frac{\t}{2\kappa(n-1)}\Big]  \Big\}  \ .
\eeqa
% where one should remember there is a convenient redefinition $\s\leftrightarrow \t$. 
The modulus of the acceleration of the end point of the open string (quark) is given by
\beq
 a^2 = (\pa_\t^2 \Y_1)^2  +(\pa_\t^2 \Y_2)^2  = \frac{\epsilon^2}{\kappa^2}\sin^2\Big[ \frac{n\t}{4(n-1)\kappa}\Big]\ . 
\eeq
The acceleration vanishes at the points 
\beq
 \tau_m = 4\kappa \pi \frac{n-1}{n}\, m, \ \ \ \ \ \ \ \ \ \ \ \   m\in\mathbb{Z} \ .
\eeq 
These  correspond to the positions of the cusps on  the small spiky string side, i.e. 
 the cusps of the small string are seen in the T-dual picture 
  as the points where the acceleration of the end-point quark vanishes. 

Note that  if the Wilson loop  had  cusps  itself  (namely,  jumps in the velocity) then the corresponding 
  small  string would have  jumps in the position. This is not
possible, i.e. 
 one cannot obtain Wilson loops with cusps  by this T-duality transform.
 In fact,  it was already argued in \cite{am} that  the cusped Wilson loops
 are  dual 
 to open strings representing scattering of gluons  (i.e. to non gauge-invariant operators).

%%%%%%%%%%%%%%%%%%%%%%%%%
\section{T-duality  in a non-compact direction  on a   cylinder}

\def\theequation{C.\arabic{equation}}
\setcounter{equation}{0}

  The world-sheet theory corresponding to the closed string is defined on the cylinder (see  fig. 2).
 %  This is normal, but the coordinate $\T$ 
   At the same time, the  coordinates $\X_\mu$ 
 that we T-dualized  in \rf{tt},\rf{116}  are non-compact   and  that  may  be a source of concern. 
 Below we shall   review the application of T-duality  in such  a case of a non-compact 
 target space coordinate (we  shall consider only one bosonic $\X_0=\T$ coordinate).
 The fermionic T-duality  and the  full quantum map  in the case of the $AdS_5 \times S^5$ superstring theory
 is left for the future work. \foot{Let us note  also  that 
 T-duality for non-compact directions only works at lowest order in  string coupling $g_s$ \ci{bm}, namely,
  at the planar level of the dual field theory.
  It might be possible to extend it further by including 
extra degrees of freedom, something that should be done if $1/N$ corrections need to be understood.}

  The generic  world-sheet   action can be written as (we ignore  the overall factor of string tension)
\bea
&&  S =  \int d\s d\t \Big[g_{\T\T}(\pa_\t \T \pa_\t \T- \pa_\s \T \pa_\s \T) 
          + 2 g_{\T i}(\pa_\t \T \pa_\t \X^i- \pa_\s \T \pa_\s \X^i) \no\\ 
&& \ \ \ \ \    \ \ \ \ \   \ \ \ \ \    \ \ \ \ \        +\  g_{i j}(\pa_\t \X^i \pa_\t \X^j- \pa_\s \X^i \pa_\s \X^j)\Big] \ , \la{c1} 
\eea
where in  our present case  the metric depends only on the coordinate $Z$ that is not displayed here. 
%Now a vector field is introduced by replacing $\pa_\s \T \rightarrow A_\s$ and
%$\pa_\t \T =  A_\t$ plus a Lagrange multiplier term enforcing the curl of $A$ to vanish:
Gauging as usual  the shifts  in $\T$   we get 
\beqa
  S &=&  \int d\s d\t \Big[g_{\T\T}(A_\t A_\t- A_\s A_\s) 
           + 2 g_{\T i}(A_\t \pa_\t \X^i- A_\s \pa_\s \X^i)
           + g_{i j}(\pa_\t \X^i \pa_\t \X^j- \pa_\s \X^i \pa_\s \X^j) \no  \\
    & &\ \ \ \ \ \ \ \ \  \ \ \ \ \ \ \ \   +\ \tilde{\T} (\pa_\s A_\t - \pa_\t A_\s)\Big] \ . \la{c2}
 \eeqa 
The path integral over  the Lagrange multiplier 
$\tilde{\T}$ enforces $A_{\s,\t}$ to be a pure gauge  and therefore the path integral over $\tilde{\T}, A_\s, A_\t$  is equivalent to the path integral over $\T$. On the other hand, integrating first over $A_\s$, $A_\t$ gives the T-dual action for $\tilde{\T}$. 

To deal with the subtetly of the theory  defined on  the cylinder  let us expand all the variables as periodic functions of $\s$
(here  assumed  to  be  $2\pi$ periodic)  
\beqa
 \T &=& \xi_0(\t)+ \sum_{n\neq 0} e^{in\s} \xi_n(\t) \ ,  \ \ \ \ \ \ \ \ \ \ \ \ \ \ \ 
  \tilde{\T} = \tilde{\xi}_0(\t)+ \sum_{n\neq 0} e^{in\s} \tilde{\xi}_n(\t) \ , \la{c3}  \\
A_\s &=& \xi_{\s0}(\t)+ \sum_{n\neq 0} e^{in\s} \xi_{\s n}(\t) \ , \ \ \ \ \ \ \ \ \ \ \ \ \ 
A_\t =  \xi_{\t0}(\t)+ \sum_{n\neq 0} e^{in\s} \xi_{\t n}(\t)   \la{c4}\ . 
\eeqa
 Ignoring first the zero modes, let us  check that the path integral over $\xi_{\s n}(\tau)$, $\xi_{\t n}(\tau)$ 
 and $\tilde{\xi}_n(\tau)$ is equivalent to the path integral over $\xi_n(\tau)$. Indeed,  the action has a term
\beq
 \int d\tau \sum_{n\neq 0}  \tilde{\xi}_{-n} (in \xi_{\t n} - \pa_\t \xi_{\s n})\ . 
\eeq
 The path integral over $\tilde{\xi}_n$ enforces the relation 
 $\xi_{\t n} = \frac{1}{in} \pa_\t \xi_{\s n}$ which means that the only independent  variables remaining are 
 $\xi_{\s n}$. Then defining $\xi_{n} = \frac{1}{in} \xi_{\s n}$  accomplishes the task. 
 
 The zero  modes in \rf{c3},\rf{c4}    present, however,   few problems.
  Notice that   $\xi_{\t 0}$  which should be equal to $\pa_\t \xi_0$
  does not appear in the  last term 
  \be S_A= \int d\tau  d \s\, \tilde{\T} (\pa_\s A_\t - \pa_\t A_\s) \la{cc4}  \ee 
  in  the action \rf{c2}. 
  To enforce the required   relation we may add an extra term   to the action   depending  on  a new variable $\mu_0(\t)$ 
 \beq
   S_0 =- \int d\t \ \mu_0(\t) (\xi_{\t 0} - \pa_\t \xi_0) =-{ 1 \ov 2 \pi}  \int d\t d\s\  \mu_0(\t) A_\t    +  \int d\t \ \mu_0(\t) \pa _\t \xi_0\ . \la{c5} 
 \eeq
  Performing the path integral over $\mu_0(\t)$ gives  $\xi_{\t 0} = \pa_\t \xi_0$ and the subsequent
  integral over $\xi_0$ completes the $\T$ with the zero mode. 
 Integrating  by parts in  the last term of \rf{c5}  gives 
\beq
   S_0 =- { 1 \ov 2 \pi}   \int d\t d\s\  \mu_0(\t) A_\t -  \int d\t \ \pa_\t \mu_0(\t)  \xi_0(\t)  +  \mu_0(\t_f) \xi_0(\t_f) - \mu_0(\t_i) \xi_0(\t_i)\ . \la{c7} 
\eeq
   If we  now integrate  over $\xi_0$  we get  $\mu_0=$const. 
  Adding  together  the extra $S_A$ and $S_0$  terms and after some integral by parts, 
we have 
\beq
 S_A +S_0 = - \int d\t d\s\ \Big[ (\pa_\s \tilde{\T}+\bar  \mu_0 )  A_\t - \pa_\t \tilde{\T} A_\s\Big]  +2 \pi \bar  \mu_0 [\xi_0(\t_f)-\xi_0(\t_i)]\ , 
 \ \ \ \ \ \ \ \ \ \   \bar \mu = {1 \ov 2 \pi} \mu_0 \ .
\eeq
 In the T-dual action we should replace $\pa_\s \tilde{\T} \rightarrow \pa_\s \tilde{\T}+ \bar \mu_0$.
 
 If we Fourier transform over the center of mass positions
$\xi_0(\t_f)$ and $\xi_0(\t_i)$ we get that the initial and final momenta in the  $\T$  direction 
 should be equal to $\mu_0$. Therefore, instead of integrating over $\mu_0$ we can fix $\mu_0$ to
 be equal to  the value of the corresponding conserved momentum, namely,  the energy  conjugate to $\T=\X_0$.

 Up to now we discussed  the theory defined on the cylinder with all the functions
periodic in $\s$.  However,  we   may  formally define the  T-dual coordinate $\tilde{\T}$ as
\beq
 \tilde{\T} =\bar  \mu_0 \s + \sum_{n\neq 0} \tilde{\xi}_{n}(\t) e^{in\s}  \ . \la{c8} 
\eeq
Including the linear term in $\s$ is allowed as 
  $\tilde{\T}$ enters in the action through its derivatives which are all periodic. 
  A problem,  however,   arises if
 we insert  vertices with momenta  in the direction of  $\tilde{\T}$:
 the presence of  a factor $\exp(i\tilde{E}\tilde{\T})$ will  break the periodicity in $\s$. 
 % is no longer
% periodic. 
% Otherwise the theory is completely defined on the cylinder. 

 Equivalently,  we  may  consider the coordinate $\tilde{\T}$ to be periodic with period  determined   by the energy, 
 $\tilde{\T} \equiv \tilde{\T} + 2\pi E$. 
 Physically, the period changes with the energy so it should not be considered a property of the dual space-time but
 of the configuration that we are considering. 
 The T-dual %is a periodic Wilson loop 
 string surface  is then an  open string world sheet   which is periodic  along $\tilde{\T}$.
 
  Another problem is that the zero mode part of the Lagrange multiplier term \rf{cc4} 
\beq
 S_{A0} = \int d\t\  \tilde{\xi}_0(\t) \pa_\t \xi_{\s n}(\t)
\eeq 
 sets $\pa_\t \xi_{\s n}(\t)=0$ as it should, but leaves a possible zero mode $\xi_{\t 0}$ independent of $\t$. 
 The  latter   
implies  the presence of a term $\T \sim \xi_{\t 0} \s$ which   should  not be allowed by 
the periodicity. 
This can be dealt with by setting it  to zero with a Lagrange multiplier.

 \def \bi {\bibitem}
 %%%%%%%%%%%%%%%%%%%%%%%%%%%%%%%

\end{document}